\documentclass[lettersize,journal]{IEEEtran}
\usepackage{amsmath,amssymb,amsfonts}
\usepackage{algorithmic}
\usepackage{algorithm}
\usepackage{amsthm}

\usepackage{booktabs}
\usepackage{array}
\usepackage[caption=false,font=normalsize,labelfont=sf,textfont=sf]{subfig}
\usepackage{textcomp}
\usepackage{stfloats}
\usepackage{url}
\usepackage{verbatim}
\usepackage{graphicx}
\usepackage{cite}
\usepackage{soul}
\usepackage{xcolor}
\newcommand{\INPUT}{\item[\textbf{Input:}]}

\newcommand{\OUTPUT}{\item[\textbf{Output:}]}

\begin{document}

\title{Denoising Diffusion Probabilistic Model for Radio Map Estimation in Generative Wireless Networks\\}

\author{
		\IEEEauthorblockN{Xuanhao Luo, 
            Zhizhen Li, 
            Zhiyuan Peng, 
            Mingzhe Chen,~\IEEEmembership{Member,~IEEE,}
            Yuchen Liu,~\IEEEmembership{Member,~IEEE}
 }

\thanks{X. Luo, Z. Li, Z. Peng, and Y. Liu are with the Department of Computer Science, North Carolina State University, Raleigh, NC, 27695, USA (Email: \{xluo26, zli92, zpeng22, yuchen.liu\}@ncsu.edu).}
\thanks{M. Chen is with the Department of Electrical and Computer Engineering, University of Miami, FL, 33146, USA (Email: \protect\url{ mingzhe.chen@miami.edu)}.} 
\vspace{-1.5em}
}

\markboth{Journal of \LaTeX\ Class Files,~Vol.~14, No.~8, August~2021}%
{Shell \MakeLowercase{\textit{et al.}}: A Sample Article Using IEEEtran.cls for IEEE Journals}

\IEEEpubid{0000--0000/00\$00.00~\copyright~2021 IEEE}

\maketitle

\begin{abstract}

The increasing demand for high-speed and reliable wireless networks has driven advancements in technologies such as millimeter-wave and 5G radios, which requires efficient planning and timely deployment of wireless access points. A critical tool in this process is the radio map, a graphical representation of radio-frequency signal strengths that plays a vital role in optimizing overall network performance. However, existing methods for estimating radio maps face challenges {due to the need for extensive real-world data collection or computationally intensive ray-tracing analyses, which is costly and time-consuming}.
Inspired by the success of generative AI techniques in large language models and image generation, we explore their potential applications in the realm of wireless networks.
In this work, we propose RM-Gen, a novel generative framework leveraging conditional denoising diffusion probabilistic models to synthesize radio maps using minimal and readily collected data.
We then introduce an environment-aware method for selecting critical data pieces,
enhancing the generative model's applicability and usability. 
Comprehensive evaluations demonstrate that RM-Gen achieves over 95\% accuracy in generating radio maps for networks that operate at 60 GHz and sub-6GHz frequency bands, outperforming the baseline GAN and pix2pix models. 
This approach offers a cost-effective, adaptable solution for various downstream network optimization tasks.

\end{abstract}


\section{Introduction}

As bandwidth-intensive applications like virtual reality, high-definition streaming, and real-time communications surge in popularity, there is an increasing need for faster and more reliable wireless networks. Millimeter-wave (mmWave) technology, notably within the 60 GHz frequency band, is spearheading advancements in indoor wireless communication by providing substantial benefits for compact, short-range networks. In parallel, advancements in 5G/6G technology are transforming outdoor cellular networks, enhancing data transmission rates, lowering latency, and significantly expanding network capacity. These technological strides are driving the need for meticulous network planning and the rapid deployment of sophisticated infrastructure. Within this evolving landscape, radio maps have proven to be indispensable for enhancing network performance and predictive configuration. These maps graphically depict received signal strength (RSS) across diverse locations, offering detailed insights into the coverage and intensity of radio-frequency (RF) signals within areas of interest. Such in-depth mappings are critical for identifying regions with fluctuating radio signals, thus facilitating targeted enhancement in network coverage and reliability.

\IEEEpubidadjcol
In practical scenarios, the value of radio maps is increasingly recognized for their role in analyzing wireless propagation, managing signal interference, optimizing capacity planning, and streamlining network deployment. These maps empower network operators to strategically install base stations, adjust antenna configurations, and promote network virtualization. 
Recent advancements in various application areas underscore the importance of effective resource management in wireless networks. These include predictive resource planning and beam management \cite{phillips2012survey, li2024contextual, li2024context}, the development of digital twin networks \cite{almasan2022network, yang2024optimizing, yang2023joint}, and the design of flight paths for drone-assisted communication systems \cite{li2019closed, kumar2019analysis}. Radio maps, by providing granular RSS data across spatial domains, enable a more equitable distribution of network resources, ensuring consistent and fair service quality across both well-served and under-served areas.
In the realm of network digitalization \cite{almasan2022network}, radio maps are instrumental in generating precise digital twins of network environments. These digital replicas, rich in location-aware RF data, are invaluable for modeling and predicting network behavior under a variety of hypothetical scenarios and channel conditions.
Moreover, the role of radio maps becomes even more critical in the context of non-terrestrial network studies, particularly in networks that incorporate unmanned aerial vehicles (UAVs) as mobile access points \cite{li2019closed}. In these scenarios, radio maps are crucial for crafting optimal UAV flight paths, thereby enhancing coverage in regions that lack robust ground-based infrastructure for a broader network accessibility.

Despite the evident advantages and widespread applications, challenges persist in generating accurate and complete radio maps. Even minor alterations in the physical environment or the relocation of transmitters can result in significant fluctuations in radio maps, as illustrated in Fig.~1(a)-(c), with the RSS measurements of three distinct access point (AP) locations in an indoor wireless local-area network (WLAN).
Consequently, collecting real-world data across all possible scenarios to obtain various radio maps is impractical due to the extensive measurement campaigns required. To address this issue, ray-tracing based analysis emerges as an alternative solution for generating radio maps within a simulation environment \cite{yun2015ray}. However, achieving an accurate radio map while considering detailed multi-path effects always demands substantial computational resources and considerable time. 
To alleviate the online computation overhead, certain deep learning-based methods have been proposed as alternative solutions~\cite{li2024map, liu2022environment}. However, these methods typically rely on extensive environmental information, including detailed geographic maps and object parameters, which may not always be readily  available. 
These challenges highlight the necessity for a cost-effective approach to derive precise and adaptable radio maps with minimal prior knowledge, which is the focus of our research. 

\begin{figure}[htbp]
\centerline{\includegraphics[width=1\linewidth]{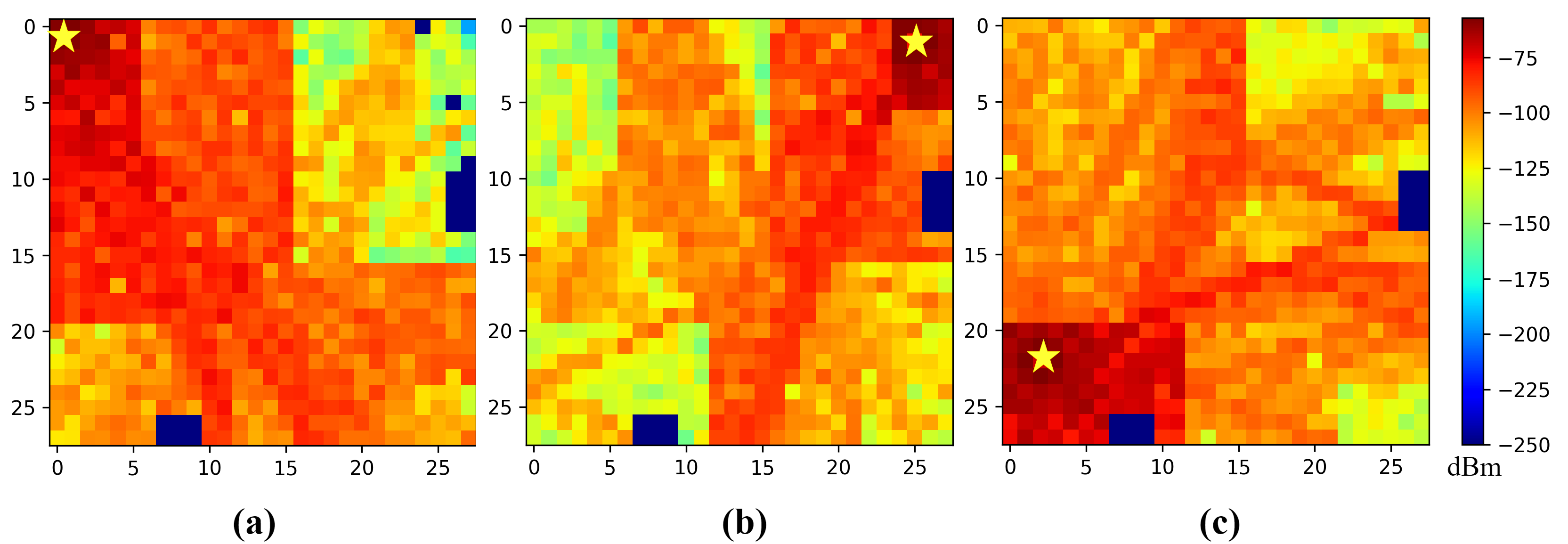}}
\caption{Radio maps with different base station (BS)/AP locations (denoted by stars) in a mmWave network scenario.}
\label{fig}
\end{figure}

Inspired by the remarkable success of generative artificial intelligence (AI) techniques, particularly the effectiveness of diffusion probabilistic models in generating realistic images, this work is motivated to explore their potential for synthesizing radio maps in wireless networks. 
The core idea of these models is to transform data from its original state into a state of pure noise through a controlled random process, and then to reverse this process to reconstruct the original RF data. The reverse process is achieved by training a neural network that learns to gradually remove noise and restore the RF data. By learning this step-by-step denoising process, diffusion models can generate new data that closely resembles the training data, showcasing excellent performance in generative tasks. Considering the capabilities of diffusion models to reconstruct data that resembles the original samples, we aim to leverage this underlying mechanism in wireless networks to generate precise radio maps using minimal and readily available data. 
In order to broaden the applicability and enhance the usability of our generative model, we also develop an environment-aware method for selecting RSS data fragments, automating the sparse data measurement process without requiring extensive collection or human intervention.
To ensure the training of our generative diffusion-based model for an adaptable map-based RSS estimation, a ray-tracing-based method is employed to synthetically collect training data covering a wide range of fine-grained network scenarios across both 60 GHz and sub-6GHz frequency bands. In summary, this work presents the following contributions:
\begin{itemize}
    \item We propose \textit{RM-Gen}, a conditional diffusion model-based framework, for generating radio maps across diverse wireless scenarios, including mmWave WLANs and 5G cellular networks, using sparse measurement data and easily obtainable environment information. To the best of our knowledge, this is the first endeavor to leverage generative diffusion models for constructing radio maps in wireless networks.
    
    \item We demonstrate the feasibility of using two accessible information pieces as input conditions for the generative model: 1) a limited amount of RSS information and/or 2) transmitter locations. This approach facilitates the cost-effective generation of radio maps, especially useful in complex scenarios where obtaining comprehensive measurements is challenging.

    \item We develop an environment-aware method for selecting critical RSS fragments without prior knowledge of scenario configurations. Additionally, we develop a data collection method to synthetically generate high-quality training data covering a wide range of fine-grained wireless scenarios, which is then used to train the generative diffusion-based model for adaptable radio map prediction.

    \item We perform comprehensive evaluations using the two collected radio map datasets across both 60 GHz and sub-6GHz frequency bands. The results show that \textit{RM-Gen} exhibits the capability to efficiently generate precise radio maps, achieving accuracy of over 95\% in both indoor and outdoor wireless network scenarios, which consistently surpasses the baseline models across various environments. 

\end{itemize}

The rest of this paper is organized as follows: Section II reviews related works in the field. In Section III, we outline the problem formulation and the preliminaries on denoising diffusion probabilistic models. Section IV presents our conditional diffusion model for radio maps generation, along with an environment-aware data selection method and a map-based data collection approach using ray-tracing analysis. Performance evaluation of our developed model, including extensive experiments and comparison with a baseline model, is presented in Section V. Section VI concludes the paper.

\section{Related Works}
\subsection{Radio Map Prediction}

Radio map prediction has been the subject of recent research efforts, with approaches ranging from interpolation-based methods like Kriging \cite{sato2017kriging, romero2020aerial}, matrix completion \cite{sun2022propagation, balachandrasekaran2021reducing}, to dictionary learning \cite{kim2013cognitive}. However, these approaches often lack the ability to learn from experience, resulting in limited generalization capabilities, especially in dynamic and evolving network environments. In contrast, recent studies have explored the application of deep learning models for radio map prediction and propagation modeling in wireless networks. For instance, U-Net, a convolutional neural network (CNN), has been employed in \cite{krijestorac2021spatial} to predict RSS by leveraging 3D city maps and signal strength samples collected from the prediction area.
\cite{teganya2021deep} proposes a data-driven approach for estimating radio occupancy maps by learning the spatial structure of propagation phenomena from past environmental measurements using completion autoencoders. In~\cite{gupta2022machine}, a CNN-based Autoencoder (AE) for feature extraction from 3D building data is employed to utilize the extracted street clutter information for path loss estimation, while \cite{hehn2023transformer} introduces an alternative transformer-based approach that incorporates map data overlaid with transmitter and receiver locations and their distances. Despite their effectiveness, these deep learning based models often require large datasets for training and rely heavily on detailed environment information for an accurate radio map prediction.
{ Recently, some studies proposed leveraging transfer learning to reduce the amount of data required for retraining radio map prediction models. 
\cite{levie2021radiounet} introduces a U-Net-based approach for radio map estimation, where transfer learning is utilized to adapt from simulated datasets to real-world scenarios using sparse measurements, thereby reducing the need for extensive data collection.
Similarly, \cite{lee2024scalable} proposes a scalable and generalizable pathloss map prediction framework that employs transfer learning to fine-tune pre-trained models on the new environments, significantly lowering the data requirements for retraining.}

With the advancement of generative AI techniques, generative adversarial networks have first emerged as a promising avenue for exploring image-based estimations. For instance, \cite{liu2020cgan} tackles radio layout design by framing it as an image-to-image translation problem, employing customized dimension-aware conditional adversarial networks (cGANs) to generate radio heatmaps and dot layouts from floor plans. Then, subsequent studies have explored different approaches to radio map estimation using such generative models, e.g., \cite{vankayala2021radio} emphasizes cost-effectiveness and computational efficiency, while \cite{kim2021access} introduces an access-point centered window to enhance accuracy and efficiency. Additionally, \cite{zhang2023rme} proposes a two-phase learning framework integrating cGANs to extract global and local features, and \cite{cisse2023irgan} devises a model to generate high-quality radio maps from floor plan images, considering diverse building materials and transmitter locations. 
{While cGAN can, in theory, be applied to generate radio maps using sparse measurement data, the aforementioned works typically rely on detailed floor plans or geographical information as input, which limits their applicability due to the time-consuming and expensive nature of gathering such dataset for every scenario.} In contrast, our research endeavors to overcome this limitation by developing a conditional diffusion model capable of generating accurate radio maps using sparse measurement data and readily available environment information. This approach offers a more practical and flexible solution for radio map generation across diverse network scenarios.

\subsection{Generative Diffusion Models for Wireless Networks}

{In recent years, diffusion models have shown remarkable success in various tasks such as image generation \cite{ho2020denoising, rombach2022high}, video generation \cite{ho2022imagen}, audio synthesis \cite{kong2020diffwave}, time series data imputation \cite{tashiro2021csdi} and forecasting \cite{li2022generative},  and spatio-temporal data prediction \cite{yang2024survey, ruhling2024dyffusion}}. Applied to wireless networks, diffusion models have been utilized for channel modeling and enhancement in wireless communication \cite{arvinte2022score, wu2023cddm, letafati2023denoising}. For instance, \cite{arvinte2022score} proposes a score-based diffusion model for channel estimation in multiple-input multiple-output (MIMO) communication systems, while \cite{wu2023cddm} introduces channel denoising diffusion models to eliminate channel noise. Unlike \cite{wu2023cddm}, which jointly trains the source-channel encoder and decoder with the diffusion model using pre-obtained channel state information (CSI), \cite{letafati2023denoising} presents denoising diffusion probabilistic models (DDPMs) for hardware-impaired communications without relying on additional decoders or CSI. 
In \cite{letafati2023generative}, the authors utilize the ``denoise-and-generate'' characteristics of DDPMs for probabilistic constellation shaping in communications. Meanwhile, \cite{kim2023learning} investigates the application of diffusion models in end-to-end (E2E) channel coding for wireless communication. 
Moreover, in \cite{liu2023deep}, a comprehensive overview of deep generation models (DGMs) and their applications in wireless networks is provided. The work proposes a DGM-empowered framework for wireless network management and conducts a case study on using diffusion models for contract generation in mobile artificial intelligence generated content (AIGC) services.

Beyond its applications in channel coding analysis, diffusion models are also being considered for image-based communications, as evidenced by works such as \cite{niu2023hybrid, yilmaz2023high, chen2024commin}. For instance, \cite{niu2023hybrid} proposes a novel hybrid joint source-channel coding (JSCC) method for image transmission using a diffusion model, which combines digital and semantic transmission parts to enhance both bandwidth efficiency and information reconstruction quality. Additionally, DDPMs are integrated with deep learning-based JSCC at the receiver for image transmission over noisy wireless channels in \cite{yilmaz2023high}, which mainly focuses on the perception-distortion trade-off in scenarios with finite block lengths where traditional coding methods are less effective. Moreover, \cite{chen2024commin} combines invertible neural networks and diffusion models for high-quality image reconstruction, particularly effective in challenging conditions such as low bandwidth and low signal-to-noise ratio. Despite these advancements, to the best of our knowledge, there is no existing work exploring the use of generative diffusion models for radio map generation. Motivated by the aforementioned impressive performance of diffusion models in image generation, channel coding, and image-based communication, this paper extends our prior work \cite{luo2024rm} by investigating a diffusion model-based approach for radio map estimation in wireless networks, incorporating elaborately designed condition encoders and an environment-aware RSS fragment selection mechanism.

\section{Problem Formulation and Preliminary on Denoising Diffusion Probabilistic Models}
In this section, we first formulate the radio map generation problem as an optimization problem. Then, we introduce the background knowledge on denoising diffusion probabilistic models. Important notations used in the paper can be found in Table~\ref{symbols}.

\begin{table}[h]
\centering
\caption{Notations and Definitions.}
\begin{tabular}{|p{0.18\linewidth}|p{0.72\linewidth}|}
\hline
\textbf{Symbol} & \textbf{Description} \\
\hline
$c$ & Predefined conditions consist of a set of parameters or contexts that serve as the input for the diffusion model \\
\hline
$\hat{M}$ & Estimated radio map \\
\hline
$M$ & Ground-truth radio map \\
\hline
$D(M, \hat{M})$ & Difference between the generated radio map and the ground-truth radio map\\
\hline
$\boldsymbol{\phi}$ & RSS fragments set, $\boldsymbol{\phi} = \{\phi_1, \phi_2,\dots,\phi_k\}$ \\
\hline
$\boldsymbol{Tx}$ & Transmitter locations set, $\boldsymbol{Tx} = \{Tx_1, Tx_2, \dots, Tx_n\}$\\
\hline
$q(x_0)$ & RSS data distribution \\
\hline
$\beta$ & A variance schedule that represents the noise level \\
\hline
$T$ & Number of time step for forward process and reverse process of the diffusion model \\
\hline
$e_{\theta}$ & The encoder used to extract the features of conditions $c$. \\
\hline
$p_{\theta}(x_0 | e_{\theta}(c))$ & The generative model for estimating the radio signal distribution $q(x_0)$\\
\hline
$\epsilon_\theta$ & The trainable denoising function estimating the noise in the reverse process of the diffusion model. \\
\hline
$\epsilon_\theta (x_t, t | e_{\theta}(c))$ & The estimated noise vector for the time step \( t \) under the given condition \( c \).\\
\hline
$V_i$ & Flattened one-dimensional vector of the RSS fragment \\
\hline
$\mathcal{M}(\cdot) $ &  Multilayer Perceptron (MLP)\\
\hline
$W_i$ & The weight of $i$-th layer of the MLP \\
\hline
$b_i$ & Biases of MLP \\
\hline
$\bigoplus_{i=1}^{n}$ & Concatenation of the conditions \\
\hline
$\mathcal{G}$ & Geometric map of the network scenario  \\
\hline
$S$ & Set of subareas for the network scenario \\
\hline
$ETR$ & Error Tolerance Rates\\  
\hline
$\gamma$ & Learning rate of \textit{RM-Gen} \\
\hline
$G$ & Generator of cGAN \\
\hline
$D$ & Discriminator of cGAN \\
\hline
$z$ & Noise vector for the generator of cGAN\\
\hline
$p_g(z)$ & Noise prior distribution for the generator of cGAN\\
\hline
$\lambda$ & Control the trade-off between the adversarial loss and the L1 loss for the baseline pix2pix model\\
\hline

\end{tabular}
\label{symbols}
\end{table}

\subsection{Problem Formulation}

In this work, the objective is to develop a generative model that can generate a complete radio map for an \(N \times N\) area, leveraging a set of predefined conditions (prior knowledge) \(c\). The conditions \(c\) consist of a set of parameters or contexts that serve as the input for our \textit{RM-Gen}.
These can include a small amount of measured RSS and transmitter (Tx) location, which are often easier to obtain in practice. In this way, such a map generation problem can be represented by a function \(F: c \rightarrow \mathbb{R}^{N \times N}\), which takes the conditions \(c\) as input and outputs the estimated radio map, denoted as \(\hat{M}\). 

Specifically, the collected RSS information, referred to as \textit{partial RSS fragments}, can be denoted as $\boldsymbol{\phi} = \{\phi_1, \phi_2,\dots,\phi_k\}$. In this situation, our task is to map $\boldsymbol{\phi}$ to $\hat{M}$, which can be represented by the function $F_1 : \boldsymbol{\phi} \rightarrow \mathbb{R}^{N \times N}$. On the other hand, the transmitter locations can be denoted as $\boldsymbol{Tx} = \{Tx_1, Tx_2, \dots, Tx_n\} = \{(x_1,y_1), (x_2,y_2), \dots, (x_n,y_n)\}$; therefore, the task becomes mapping $Tx$ to $\hat{M}$, represented as $F_2 : \boldsymbol{Tx} \rightarrow \mathbb{R}^{N \times N}$.

With both types of conditions described above, the objective of our problem can be transformed to minimize the difference between the generated radio map \(\hat{M}\) and the ground-truth radio map \(M\), where \(M \in \mathbb{R}^{N \times N}\) is based on empirical or measurement results. This difference can be quantified by \(D(M, \hat{M})\), resulting in the following optimization problem:
\begin{equation}
\begin{gathered}
{\min_{{\hat{M}}} D(M,\hat{M}(c))} \\
{\text { s.t. } D(M, \hat{M}(c)) = \sum_{i=1}^{N} \sum_{j=1}^{N} |M_{ij} - \hat{M}_{ij}(c)|,}
\end{gathered}
\end{equation}
where \(i = 1, 2, \ldots, N\) and \(j = 1, 2, \ldots, N\) represent the rows and columns of grid Rx within the radio map, respectively. The optimal result can be achieved by iteratively training a generative diffusion model such that \(\hat{M}\) closely aligns with \(M\) under a given set of conditions \(c\).

\subsection{Denoising Diffusion Probabilistic Models}
In general, a diffusion model consists of two processes: a forward process and a reverse process. The forward process is a Markov chain that adds Gaussian noise at each time step. Let $q(x_0)$ be the real data distribution, the forward process can be defined as $q(x_t|x_{t-1})$, where $q(x_t)$ is the noisy data at time step $t$. The Gaussian noise added at each time step $t$ is controlled by a variance schedule $\beta_1, \dots, \beta_T$, where $T$ is the total time step. Consequently, the forward process can be derived as:
\begin{equation}
q(x_{1:T}|x_0) = \prod_{t=1}^{T} q(x_t|x_{t-1}),
\end{equation}
where
\begin{equation}
q(x_t|x_{t-1}) = \mathcal{N}(x_t; \sqrt{1 - \beta_t}x_{t-1}, \beta_t \mathbf{I}).
\end{equation}
During the diffusion process, \(\beta_t \in (0, 1)\) always increases as $t$ grows, i.e., $0 < \beta_1 < \beta_2 < \dots < \beta_T < 1$. For \(T \rightarrow \infty\), RSS data \(x_T\) will eventually approach an isotropic Gaussian distribution. At time step $t$, the noisy map \( x_t \) is sampled from a conditional Gaussian distribution with a mean of \( \mu_t = \sqrt{1 - \beta_t}x_{t-1} \) and a variance of \( \sigma_t^2 = \beta_t \), hence, 
\begin{equation}
     x_t = \sqrt{1 - \beta_t}x_{t-1} + \sqrt{\beta_t}\epsilon, 
\end{equation}
where \( \epsilon \sim \mathcal{N}(0, \mathbf{I}) \). According to the property of Gaussian distribution, $x_t$ can be sampled at an arbitrary time step $t$ in a closed form, i.e.,
\begin{equation}
    q(x_t|x_0) \sim \mathcal{N}(x_t; \sqrt{\bar{\alpha}_t}x_0, (1 - \bar{\alpha}_t)\mathbf{I}),
\end{equation}
where $\alpha_t := 1-\beta_t$ and $\bar{\alpha}_t := \prod_{i=1}^t \alpha_i$. Then, $x_t$ can be further formulated as:
\begin{equation}
    x_t = \sqrt{\bar{\alpha}_t}x_0 + \sqrt{1 - \bar{\alpha}_t}\epsilon. 
\end{equation}

In the reverse process, the diffusion model recovers $\mathbf{x_0}$ by denoising $x_t$. Such an reverse process can be defined as a Markov chain as:
\begin{equation}
    p_{\theta}(x_{0:T}) := p(x_T) \prod_{t=1}^{T} p_{\theta}(x_{t-1} | x_t),
\end{equation}
where $x_T \sim \mathcal{N}(\mathbf{0, I})$, and $p_{\theta}(x_{t-1} | x_t)$ can be represented as:
\begin{equation}
    p_{\theta}(x_{t-1} | x_t) := \mathcal{N}(x_{t-1}; \mu_\theta(x_t, t), \Sigma_\theta(x_t, t)).
\end{equation}

\noindent Following the DDPM approach proposed in \cite{ho2020denoising}, it is proved that the reverse process can learn the mean value of $\mu_\theta(x_t, t)$. When setting $\Sigma_\theta(x_t, t) = \sigma_t^2\mathbf{I}$, where $\sigma_t^2 = \frac{1 - \bar\alpha_{t-1}}{1 - \bar\alpha_t} \beta_t$, $\mu_\theta(x_t, t)$ can be derived as:

\begin{equation}
\mu_\theta(x_t, t) = \frac{1}{\sqrt{\alpha_t}} \left( x_t - \frac{\beta_t}{\sqrt{1 - \bar{\alpha}_t}} \epsilon_\theta(x_t, t) \right),
\end{equation}
where $\epsilon_\theta$ is the trainable denoising function estimating the noise in the reverse process. Based on this, we can formulate the loss function of the DDPM model as:
\begin{equation}
\mathcal{L} = \underset{x_0 \sim q(x_0), \epsilon \sim \mathcal{N}(0, \mathbf{I}), t}{\mathbb{E}} \left\| \epsilon_\theta (x_t, t) - \epsilon \right\|_2^2. 
\end{equation}

\begin{figure*}[htbp]
\centerline{\includegraphics[width=0.93\linewidth]{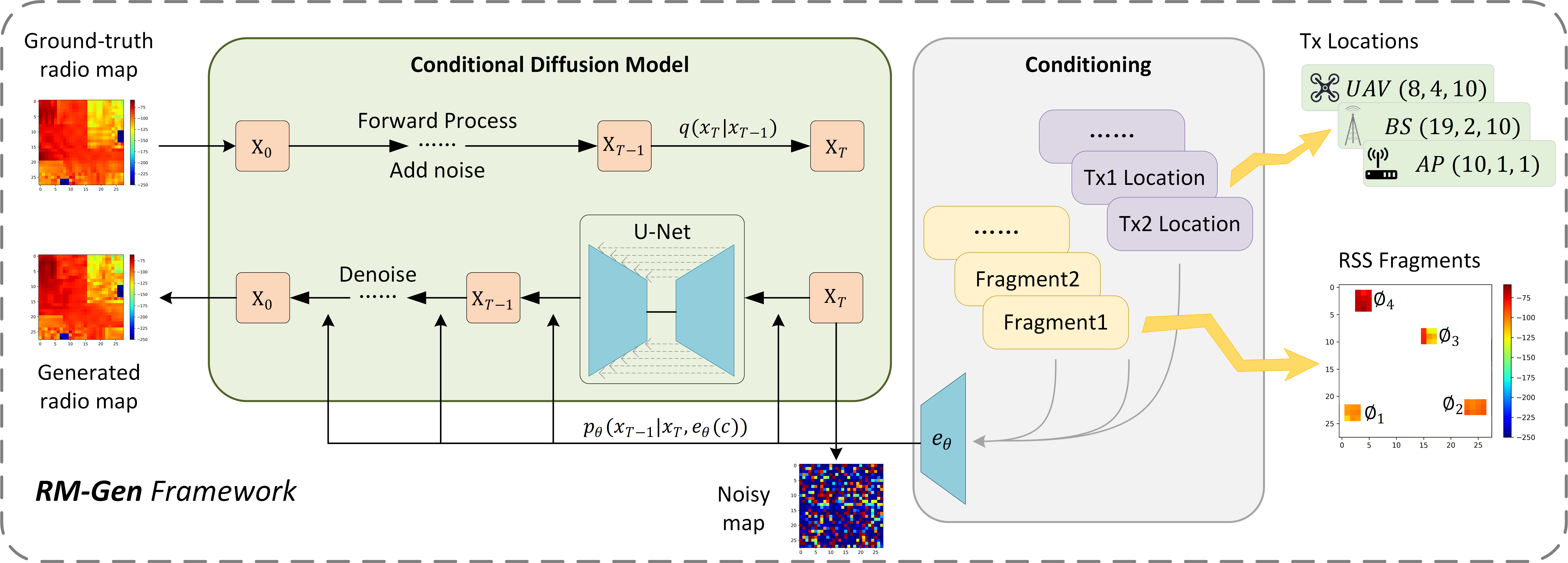}}
\caption{Overview of our conditional diffusion model \textit{RM-Gen}.}
\label{fig2_overview}
\end{figure*}

\section{Proposed Scheme for Radio Map Generation}
In this section, we first introduce our conditional denoising diffusion probabilistic model, detailing the generation process based on different input information. We then propose an environment-aware method that automates the selection of data fragments based on the complexity of the network scenario. Additionally, we introduce our map-based data collection method to train the diffusion model, leveraging RF-based ray-tracing analysis. 

\subsection{Conditional DDPM for Radio Map Generation}
In our \textit{RM-Gen}, the forward diffusion process is formulated with Eq.~(2)-(6) as detailed in Sec. III.B. Considering the two conditions, i.e., partial RSS fragments and Tx locations, the objective is to train a generative model $p_{\theta}(x_0 | e_{\theta}(c))$ capable of estimating the radio signal distribution $q(x_0)$, where $e_{\theta}$ represents the encoder used to extract the features of conditions $c$. As a result, the reverse process can be derived as:
\begin{equation}
    p_{\theta}(x_{0:T} | e_{\theta}(c)) := p(x_T) \prod_{t=1}^{T} p_{\theta}(x_{t-1} | x_t, e_{\theta}(c)),  x_T \sim \mathcal{N}(\mathbf{0, I}),
\end{equation}
\begin{equation}
    p_{\theta}(x_{t-1} | x_t, e_{\theta}(c)) := \mathcal{N}(x_{t-1}; \mu_\theta(x_t, t | e_{\theta}(c)), \Sigma_\theta(x_t, t | e_{\theta}(c))).
\end{equation}
As described in Sec. III.B, we have $\Sigma_\theta(x_t, t | e_{\theta}(c)) = \Sigma_\theta(x_t, t) = \sigma_t^2\mathbf{I}$, where $\sigma_t^2 = \frac{1 - \bar\alpha_{t-1}}{1 - \bar\alpha_t} \beta_t$. Hence, the diffusion model can learn the mean $\mu_\theta(x_t, t | e_{\theta}(c))$ represented as:
\begin{equation}
\mu_\theta(x_t, t | e_{\theta}(c)) = \frac{1}{\sqrt{\alpha_t}} \left( x_t - \frac{\beta_t}{\sqrt{1 - \bar{\alpha}_t}} \epsilon_\theta(x_t, t | e_{\theta}(c)) \right).
\end{equation}
Next, the reverse process of our conditional DDPM turns to train $\epsilon_\theta$ by minimizing the following loss function, {similar to Eq.~(10) but with the addition of the condition $c$, which guides the denoising process based on the external inputs like RSS fragments or Tx locations:}
\begin{equation}
\mathcal{L} = \underset{x_0 \sim q(x_0), \epsilon \sim \mathcal{N}(0, \mathbf{I}), t}{\mathbb{E}} \left\| \epsilon_\theta (x_t, t | e_{\theta}(c)) - \epsilon \right\|_2^2. 
\end{equation}

{In particular, the encoder \( e_{\theta} \) is adeptly designed to process the set of conditions. Two encoders are developed to accommodate the considered two types of inputs. The specific design of the encoders is detailed in Sec. IV.B.}

Lastly, to predict the noise vector \( \epsilon \) in the reverse process, the denoising function \( \epsilon_\theta (x_t, t | e_{\theta}(c)) \) is realized by training a U-Net~\cite{ronneberger2015u}. The U-Net is a convolutional neural network architecture initially conceived for biomedical image segmentation, renowned for its effectiveness in handling complex image data. Our \textit{RM-Gen} employs a U-Net to first downsample the input noisy map, effectively reducing its dimensionality while preserving essential features of RSS distribution. This is achieved through a series of convolutional and pooling layers, which systematically reduce the spatial resolution of the input map. Following this, the subsequent component of the U-Net is strategically crafted to process and abstract the most salient information from the input map. This serves as a filter, enabling the retention of only the most critical spatial features of the RSS data needed for precise noise prediction. After the downsampling phase, the U-Net transitions into an upsampling process aimed at restoring the dimensionality and refining the features of the input data. Through a series of up-convolutional layers, the network progressively reconstructs the higher-resolution map features, integrating the learned context from the downsampling path. It is during this upsampling phase that the U-Net predicts the noise vector \( \epsilon_\theta (x_t, t | e_{\theta}(c)) \) for the specific time step \( t \) under the given condition \( c \). The detailed diffusion model framework of our \textit{RM-Gen} is shown in Fig.~\ref{fig2_overview}. The training process and sampling process are summarized in Algorithm \ref{train} and Algorithm \ref{sample}, respectively.

\begin{algorithm}
    \caption{\textit{RM-Gen} Training}
    \label{train}
    \begin{algorithmic}[1]
        \INPUT Radio map dataset following $x_0 \sim q(x_0)$ with condition $\boldsymbol{c}$
        \OUTPUT Trained conditional DDPM model
        \REPEAT
            \STATE Sample $x_0 \sim q(x_0)$ with condition $\boldsymbol{c}$;
            \STATE Sample diffusion step $t \in$ Uniform $(1, \dots,T)$;
            \STATE Sample noise $\epsilon \sim \mathcal{N}(0, \mathbf{I})$;
            \STATE Take gradient descent step on $\left\| \epsilon_\theta (x_t, t | e_{\theta}(c)) - \epsilon \right\|_2^2$;
        \UNTIL {converged}
    \end{algorithmic}
\end{algorithm}

\begin{algorithm}
    \caption{\textit{RM-Gen} Sampling}
    \label{sample}
    \begin{algorithmic}[1]
        \INPUT Trained conditional DDPM model, condition $\boldsymbol{c}$
        \OUTPUT Generated radio map $x_0$

        \STATE Sample $x_T \sim \mathcal{N}(0, \mathbf{I})$;
        \FOR{$t=T,\dots,1$}

            \STATE Sample $z \sim \mathcal{N}(0, \mathbf{I})$ if $t > 1$, else $z = 0$;
            \STATE $x_{t-1} = \frac{1}{\sqrt{\alpha_t}} \left( x_t - \frac{\beta_t}{\sqrt{1 - \bar{\alpha}_t}} \epsilon_\theta(x_t, t | e_{\theta}(c)) \right) + \sqrt{1 - \alpha_t}z$;
        \ENDFOR
        \RETURN{$x_0$}

    \end{algorithmic}
\end{algorithm}

\subsection{{Condition Encoders Design}}

{Regarding the partial RSS fragment, denoted as $\boldsymbol{\phi} = \{\phi_1, \phi_2,\dots,\phi_k\}$, we first perform a flattening operation on each fragment, converting the matrix that represents a segment of the radio map into a one-dimensional vector. Specifically, for each fragment, we flatten the matrix into a one-dimensional vector, which is represented as:}
\begin{equation}
Flatten(\phi_i) = V_i.
\end{equation}
This flatten operation transforms the two-dimensional matrix \( \phi_i \) into a one-dimensional vector \( V_i \). {Without this transformation, the model would not be able to handle the multi-dimensional data.} {Once each RSS fragment is flattened, we concatenate these vectors to form a single, comprehensive input vector that represents the entire set of fragments.} Mathematically, this can be represented as:
\begin{equation}
V = V_1 \oplus V_2 \oplus \cdots \oplus V_n,
\end{equation}
where \( \oplus \) signifies the concatenation operation. The resultant vector \( V \) embodies a linear arrangement of all the RSS fragments, thereby serving as a consolidated input for subsequent neural network processing. { To handle cases where the number of fragments varies, a padding mechanism can be employed. For instance, if the number of RSS fragments is less than the maximum expected input size, the vector $V$ can be padded with zeros to maintain a consistent input dimension.} {This step ensures that the model has access to the entire set of relevant features from the fragments, allowing it to capture the relationships between different parts of the radio map and generate accurate results.}

After performing the flatten and concatenate operations, we implement a three-layer Multilayer Perceptron (MLP), which consists of three 
linear layer with the non-linear activation functions,   to process these concatenated vectors as follow:
\begin{equation}
    \mathcal{M}(V) = f(W_3 \cdot (f(W_2 \cdot (f(W_1 \cdot c + b_1)) + b_2)) + b_3), 
\end{equation}
where \( f \) is an activation function, and \( W_1, W_2, W_3 \) along with \( b_1, b_2, b_3 \) represent the weights and biases of each layer, respectively. Overall, the encoder \( e_{\theta}(c) \) for the partial RSS fragments consists of flattening, concatenation, and subsequent processing through the MLP, which can be summarized as:
\begin{equation}
e_{\theta}(c) = \mathcal{M}\left(\bigoplus_{i=1}^{n} Flatten(\phi_i)\right).
\end{equation}
To be specific, \( \bigoplus_{i=1}^{n} Flatten(\phi_i) \) represents the concatenation of the flattened vectors of each fragment \( \phi_i \), where each \( \phi_i \) is transformed into a one-dimensional vector \( V_i \) through the flatten operation. The resultant concatenated vector is then fed into a MLP, denoted by \( \mathcal{M}(\cdot) \), which processes the concatenated vector to produce the final encoded representation. For the selection of multiple RSS fragments,  we propose collecting fragments from relatively \textit{dispersed} locations, enabling the capture of RSS information across a broader range of positions for more accurate radio map generation.

Similarly, when utilizing Tx locations as the conditional input to \textit{RM-Gen}, the set of conditions can be represented as the combination of coordinates \( c = \{Tx_1, Tx_2, \ldots, Tx_n\} \). For each BS/AP location \( Tx_i \), an embedding process is performed, and then the dedicated encoder concatenates these individual embeddings to form a comprehensive representation as follow:
\begin{equation}
e_{\theta}(c) = \bigoplus_{i=1}^{n} Embedding(Tx_i),
\end{equation}
where \( \bigoplus \) denotes the concatenation of the embeddings of each \( Tx_i \). Employing distinct encoders for various conditions allows the network to effectively encode relevant features based on the diverse input conditions, thereby enhancing its adaptive performance in generating radio maps.

{ These two encoders are designed to process the two types of readily available conditions independently. When both conditions need to be used simultaneously, a straightforward approach is to process each condition with its respective encoder and then combine the encoded outputs using a shared module, such as an MLP or attention mechanism, before passing them into the generation model.}

\subsection{Environment-Aware RSS Fragments Selection}

Generating accurate radio maps using our DDPM requires the strategic selection of RSS fragments as the condition $c$, which are instrumental in the training and sampling process. The selection of the RSS fragments can also significantly influences the performance of our radio map generator. Typically, extensive RSS fragments contain rich information critical for accurately modeling signal propagation, particularly in complex network scenarios characterized by dense obstructions like outdoor buildings or indoor furniture. 
However, selecting high-quality, critical RSS fragments poses inherently challenging, often requiring specialized knowledge to analyze detailed network environment information, such as wall materials, signal frequencies, and potential sources of interference. 
To address this complexity, we propose an environment-aware RSS fragment selection method based on the distribution of obstacle density in the scenario, which simplifies the task of generating accurate radio maps without requiring an in-depth understanding of the network's physical and technical specifics.

\begin{algorithm}
    \caption{Environment-Aware RSS Fragment Selection}
    \label{complexity}
    \begin{algorithmic}[1]
        \INPUT Geometric map of the network scenario $\mathcal{G}$, number of subareas $n$, number of fragments $m$
        \OUTPUT Selected fragments set $\boldsymbol{\phi}$
        \STATE Initialize an empty list $S$;
        \STATE Initialize an empty list $\phi$;
        \STATE Divide $\mathcal{G}$ into $i$ uniformly sized subareas $\{S_1, S_2, \dots, S_n\}$ and record to \textbf{$S$};
        \STATE $i=0$;
        \WHILE{$i \leq n$}
            \STATE Count the density of obstacles in $S_i$;
            \STATE $i=i+1$;
        \ENDWHILE
        \STATE Sort \textbf{$S$} in descending order based on the density of obstacles;
        \STATE $j=0$;
        \WHILE{$j \leq m$}
            \STATE $\phi_j \leftarrow $ the RSS fragment centered at $S_j$;
            \STATE $\phi=\phi\cup\{\phi_j\}$;
            \STATE $j=j+1$;
        \ENDWHILE
        \STATE \textbf{return} $\phi$
    \end{algorithmic}
\end{algorithm}

As summarized in Algorithm~\ref{complexity}, the RSS fragment selection process begins by initializing two lists: $\phi$, to store selected RSS fragment candidates, and $S$, to include subareas (Lines 1, 2), where the geometric map of the network scenario $\mathcal{G}$ is divided into uniformly sized subareas in $S$ (Line 3). Specifically, the complexity of each subarea is evaluated based on its regional obstacle density (Lines 4-8). The subareas with higher obstacle densities are presumed to have more significant impacts on signal characteristics due to multi-path effects, making them prime candidates for RSS fragment selection. Such density measurement involves counting the number and occupied area proportion of obstacles within each subarea, reflecting the potential for altering signal reflection and diffraction patterns. Next, these subareas are sorted in descending order based on their obstacle densities to prioritize areas with higher complexity for RSS fragment selection (Line 9). Starting from the highest complexity subarea, we then select the pop-outed RSS fragments until we reach the required number of fragments, denoted as $m$. Each selected RSS fragment, $\phi_j$, is centered in one of the sorted subareas (Lines 10-15). The output of this algorithm is a set of RSS fragments, $\phi$, that are likely to contain the representative information for generating accurate radio maps (Line 16). 

It is worth noting that the proposed environment-aware method streamlines the data selection process, eliminating the need for specialized knowledge or detailed environmental measurements, making it accessible and practical for operators regardless of their expertise level. 
Another straightforward approach is human-labeled selection, which involves a thorough analysis of the network environment's characteristics and signal propagation. Despite requiring extensive offline labeling effort, this method is expected to achieve optimal results of fragment selection. We will also use this \textit{Labeled} method as a comparison point in Sec. V.B.

\subsection{Map-based Data Collection with Ray-tracing Analysis}

To augment the map-based RSS dataset for training our diffusion model, we utilize the commercial ray tracer \textit{Wireless Insite}$^\circledR$ \footnote{https://www.remcom.com/wireless-insite-em-propagation-software} to synthetically generate radio map datasets in both indoor mmWave and an outdoor sub-6GHz network scenarios. Specifically, for the mmWave dataset, we focus on the 60 GHz frequency band, known for its high bandwidth and unique propagation characteristics in WLAN environments. In this context, arbitrary 3-D indoor scenarios can be configured, such as an office/lab environment layout with dimensions of 14m $\times$ 14m $\times$ 3m as depicted in Fig.~3(a). These indoor scenarios can include various objects such as wooden chairs, glass tables, and wooden cabinets. The AP serves as the transmitter and can be deployed at arbitrary location to emit signals. 
To obtain the entire radio map, we meticulously divide the space into a number of small grids and deploy a client receiver (Rx) every 0.5m to capture the intricate details of 60 GHz signal propagation and attenuation. Next, the ray-tracing analysis simulates the propagation of electromagnetic waves by tracing the paths of individual rays as they interact with surfaces and objects within the environment. Finally, the received signal strength at each Rx is calculated based on the contributions of all traced rays, accounting for their intensities, phases, and arrival times. This process allows us to effectively collect $\sim$ 30,000 radio maps with different scenario configurations, offering invaluable insights into the behavior of 60 GHz mmWave communications within a controlled setting.

\begin{figure}[t]
\centerline{\includegraphics[width=0.9\linewidth]{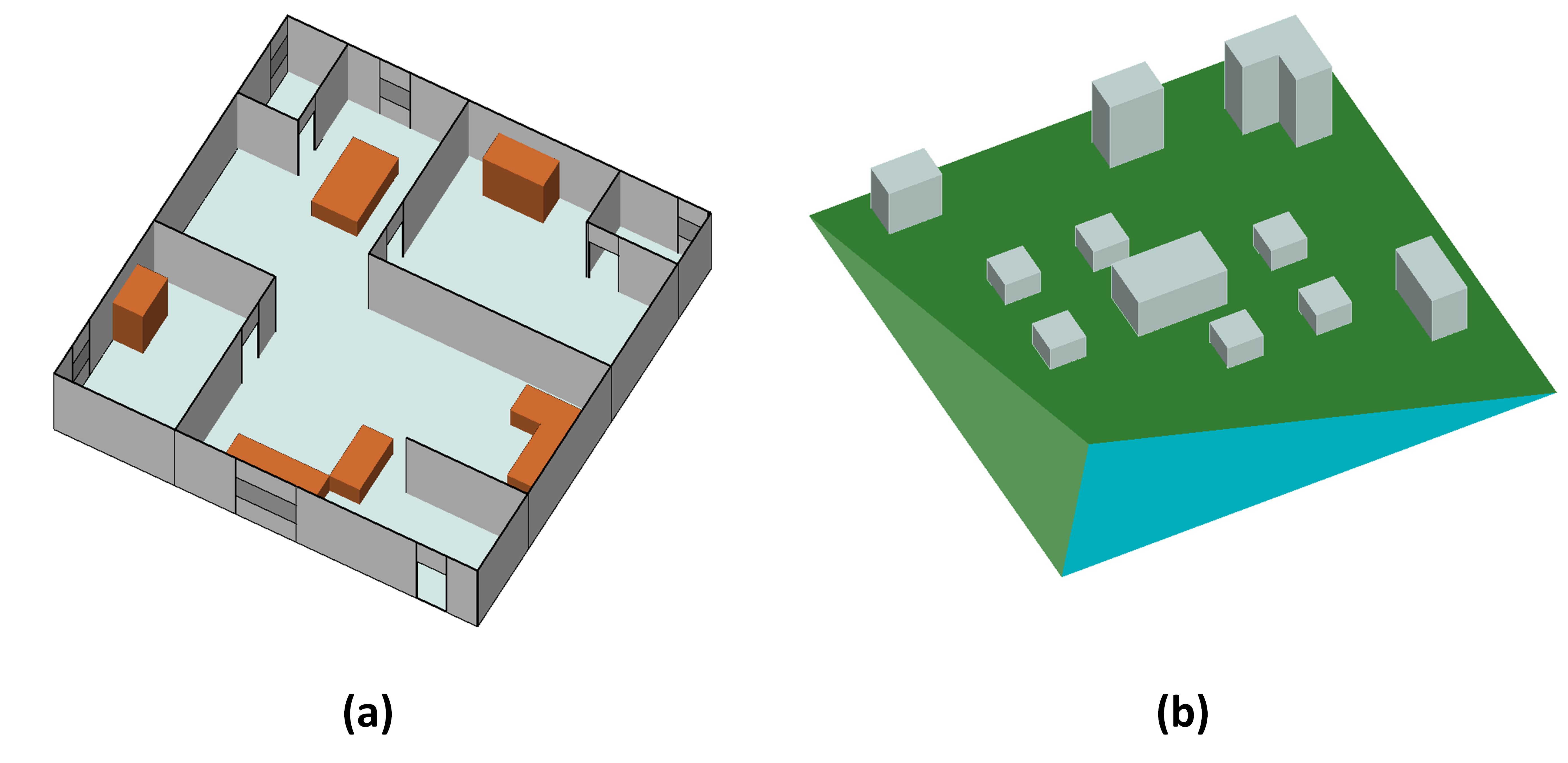}}
\caption{3-D scenario layout of (a) an indoor scenario and (b) an outdoor scenario.}
\label{fig3}
\end{figure}

For the sub-6GHz network dataset, we construct 3-D square outdoor open scenarios, as depicted in Fig. 3(b), which include features such as the ocean (blue region), the beach (light green region), the land (dark green region), and various buildings of different sizes and heights. Similar to the mmWave data collection, BSs operating at a frequency of 3.7 GHz can be randomly deployed in the scenario. This frequency band is chosen to align with standard 5G frequency bands and to ensure realistic simulation of outdoor cellular networks. Considering the longer-range propagation of sub-6GHz signals, the grid-based Rx are deployed at intervals of every 10 meters within the environment. Overall, approximately 26,000 radio maps are collected in such outdoor network scenarios. For simplicity, we denote the above two datasets as \textbf{RM-In} and \textbf{RM-Out}, respectively. It is worth noting that the ray-tracing technique is often used to estimate ground-truth data on RSS for different locations in each data set instance.  Measurement studies have demonstrated that the signal profiles produced by ray-tracing techniques are quite close to real measurements in various wireless scenarios~\cite{wang20206g, neekzad2007comparison}.

\section{Performance Evaluation}
To evaluate the effectiveness of our proposed \textit{RM-Gen}, we conduct extensive evaluations using the two collected datasets. We begin by demonstrating the efficacy of our core generative module, DDPM. Moreover, we evaluate the performance of \textit{RM-Gen} using partial RSS fragments and Tx locations as the input condition, respectively. We also compare the performance of our proposed \textit{RM-Gen} with a baseline model, conditional generative adversarial network (cGAN). Lastly, we demonstrate some practical use cases of \textit{RM-Gen}.

\subsection{Performance of DDPM}

\begin{figure}[htbp]
\centerline{\includegraphics[width=1\linewidth]{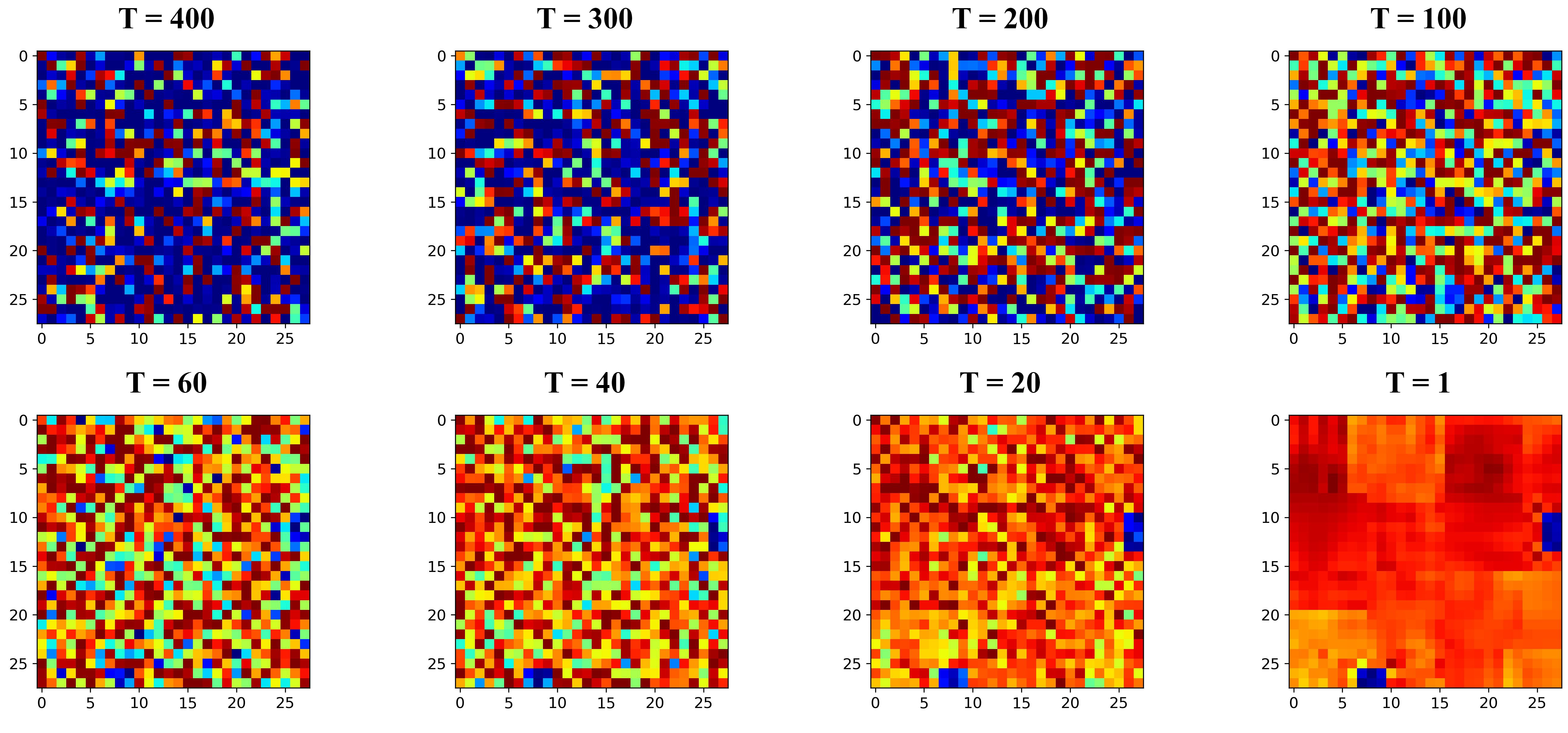}}
\caption{Map generation process over $T$ steps.}
\label{fig4c}
\end{figure}

\begin{figure}[htbp]
\centerline{\includegraphics[width=1\linewidth]{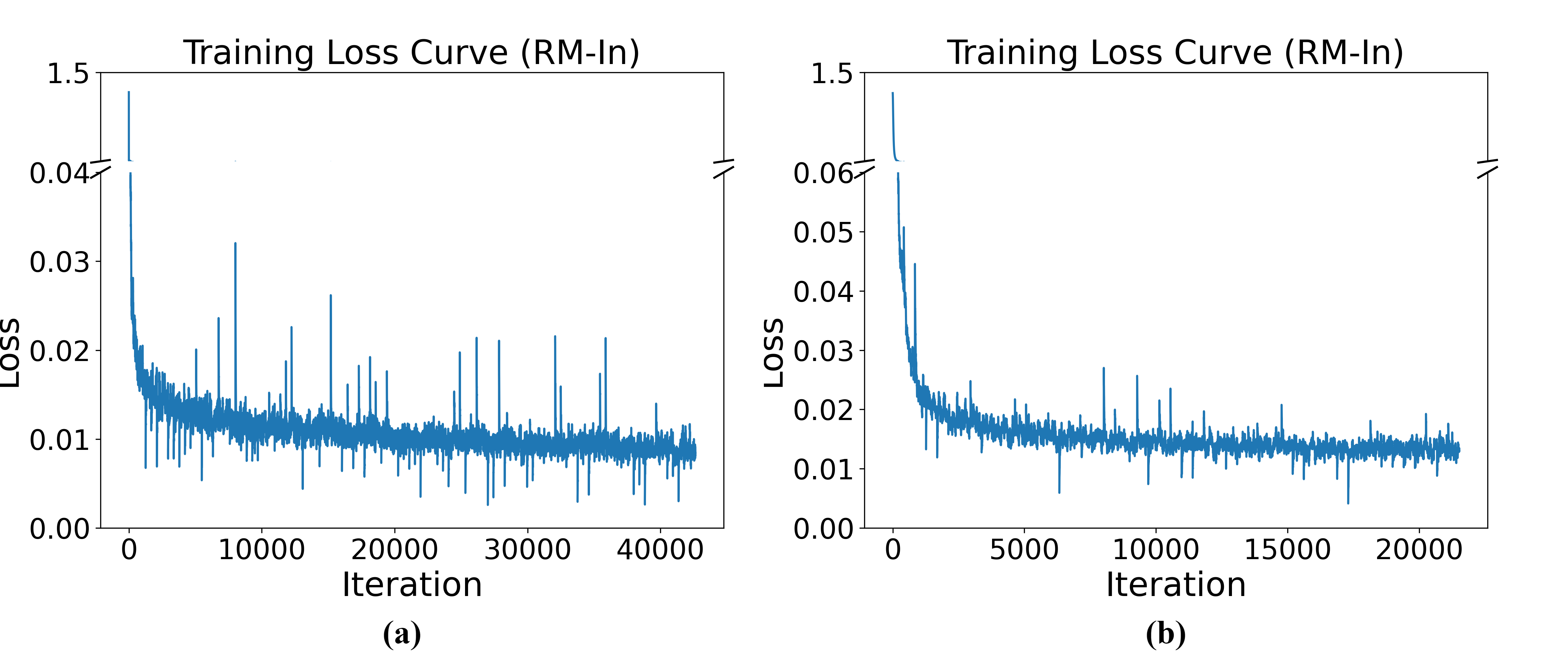}}
\caption{Model loss curve using (a) partial RSS fragments and (b) Tx locations as conditions.}
\label{fig4a}
\end{figure}

We divide the experimental settings into two parts. In the first part,  partial RSS fragments are used as conditions to generate full radio maps, and in the second part, we use Tx locations as conditions for training DDPM. As describe in Sec. IV, the forward process adds noise to the initial map according to the variance schedule $\beta$ over $T$ time steps. For our evaluations, we set $T = 400$ time steps, and the variance increases linearly from $\beta_1 = 10^{-4}$ to $\beta_T = 0.02$. We randomly select a radio map from the RM-In dataset and record  its generation process over $T$ steps as shown in Fig.~\ref{fig4c}. As can be seen from Fig.~\ref{fig4c}, the radio map is gradually denoised from the noisy map. To train the diffusion model generating radio maps using the measured RSS data pieces, we set a learning rate $\gamma = 10^{-4}$ and employ Adaptive Moment Estimation (Adam) to adjust $\gamma$ for a total of 100 epochs. The training loss curve for this setup on the RM-In dataset, depicted in Fig.~\ref{fig4a}(a), demonstrates a consistent decrease in loss value over iterations, indicating an improvement in the model's accuracy during noise prediction.

In the second part, we utilize Tx locations as conditions to generate radio maps. We set the learning rate $\gamma = 10^{-5}$ and use Adam to adjust $\gamma$ across 50 epochs. Fig.~\ref{fig4a}(b) illustrates the training loss curve for this scenario on the RM-In dataset, where a similar trend of decreasing loss values suggests the model’s noise prediction capability.  
{It is important to note that the loss in diffusion models does not directly compare the generated radio map to the ground truth but instead measures the difference between the predicted noise and the actual noise added during the forward diffusion process. Therefore, the final loss values provide an indication of the accuracy trend, though not a direct measure of map precision.}
Comparing the loss curves in Fig.~\ref{fig4a}(a) and Fig.~\ref{fig4a}(b), it can be observed that the performance using Tx locations as the condition tends to converge around 0.02, whereas the curve with partial RSS fragments as the condition converges closer to 0.01. 
{The richer, more detailed information provided by RSS fragments allows the model to better estimate the added noise during the forward diffusion process. This improves the DDPM's ability to reverse the process and generate more accurate radio maps. In contrast, Tx locations provide only positional coordinates that, while useful, do not contain as much information about signal propagation, making noise prediction less precise.}
This observation aligns with subsequent evaluation results, indicating that the DDPM trained with partial RSS fragments are more adept at predicting noise and therefore, it can generate more accurate radio maps. For the RM-Out dataset, the trend of loss curve for the two experiments is similar to the loss curve on RM-In dataset, which is omitted here due to space constraints.

\subsection{Radio Map Generation with RSS Fragments}

To demonstrate the effectiveness using \textit{RM-Gen} in generating radio maps, we first evaluate the model performance using partial RSS fragments as the prior knowledge, which then gradually compensates for the missing segments to form a complete radio map. 
In this work, we adopt the Error Tolerance Rates (ETRs) to assess the accuracy of our generated radio maps.
The ETR is a metric that specifies the maximum allowable percentage difference between the RSS values in a generated radio map and those in a ground-truth radio map. For instance, an ETR of 0.10 implies that a performance difference ratio of 10\% is acceptable, and a lower ETR indicates more stringent criteria for the quality of the radio map.
We conduct experiments with varying percentages of known RSS fragments, ranging from 5\% to 15\%, to examine their impact on the accuracy of the generated radio maps.

\subsubsection{Evaluation of data fragment selection}
\begin{figure}[t]
\centerline{\includegraphics[width=1\linewidth]{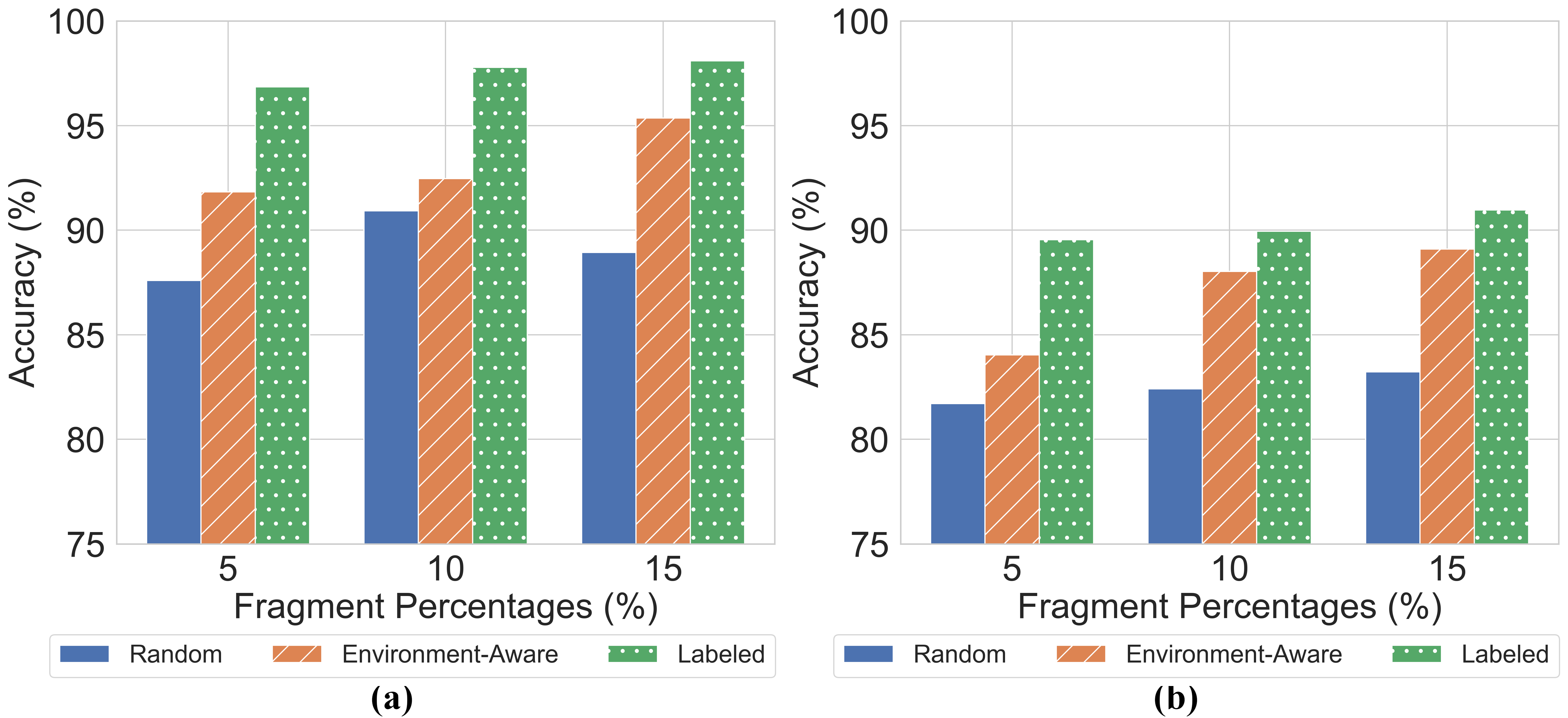}}
\caption{Generation accuracy with different RSS selection methods for (a) indoor mmWave WLANs and (b) sub-6GHz outdoor networks.}
\label{fig_selection}
\end{figure}

First, we evaluate the varying effectiveness of different RSS fragment selection methods on the accuracy of radio maps generated by \textit{RM-Gen}, with ETR = 0.10. As expected, it is evident from Fig.~\ref{fig_selection} that the \textit{Labeled} method, which relies on human-selected, strategically-placed fragments, significantly outperforms the other methods across both datasets and various fragment percentages. It achieves the highest accuracy, particularly at higher data fragment percentages, reaching up to 98.09\% on the RM-In dataset and 90.97\% on the RM-Out dataset. This suggests that having expert-driven selection of RSS fragments greatly enhances the model's ability to closely replicate the ground-truth radio maps. 

On the other hand, the \textit{Random} method, which arbitrarily selects a specific portion of fragments without any strategic consideration, consistently shows the lowest performance. This indicates its limited capability in capturing the essential features necessary for an accurate radio representation. Meanwhile, our proposed \textit{Environment-Aware} method can significantly outperforms the \textit{Random }method, especially as more data fragments are used. For instance, with 15\% fragments, the \textit{Environment-Aware} method reaches an accuracy of 95.36\% on the RM-In dataset and 89.10\% on the RM-Out dataset, demonstrating its effectiveness as a viable alternative when detailed environmental information or expert knowledge is not available. Despite the offline labeling efforts required, the subsequent experiments will continue to leverage RSS fragments selected through the \textit{Labeled} method, given its superior results from Fig.~\ref{fig_selection}. This ensures the demonstration of our generative model's optimal performance in estimating high-fidelity radio maps.

\subsubsection{Evaluation of generative model performance}
\begin{figure}[htbp]
\centerline{\includegraphics[width=1\linewidth]{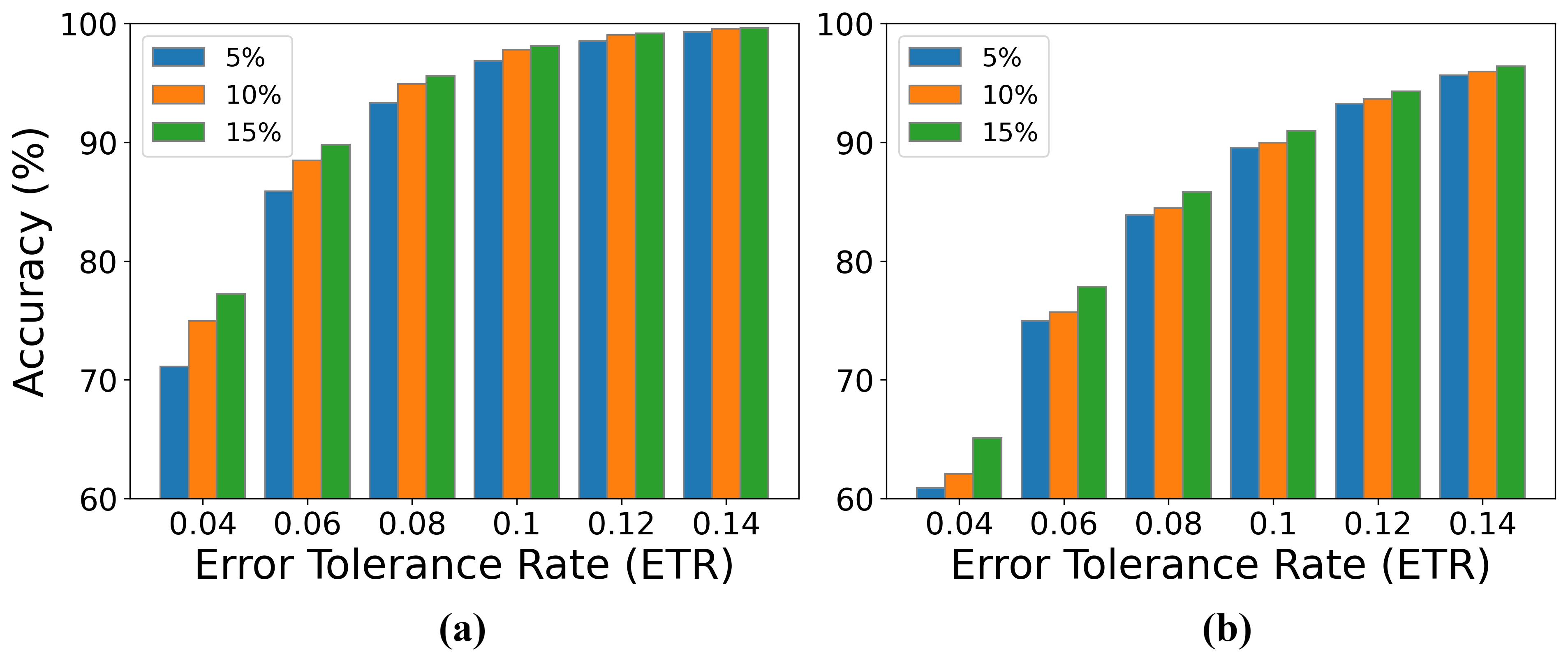}}
\caption{Generation accuracy using partial RSS fragments for (a) indoor mmWave WLANs and (b) sub-6GHz outdoor networks.}
\label{fig5a}
\end{figure}

The accuracy of generated radio maps from \textit{RM-Gen} in both indoor mmWave and outdoor sub-6GHz networks under varying ETRs are shown in Fig.~\ref{fig5a}.
In Fig.~\ref{fig5a}(a), concerning indoor mmWave scenarios, there is a noticeable trend of improved performance with increasing fragment percentages across all ETRs. This suggests that utilizing more data pieces enhances the model's capability to generate accurate radio maps. Additionally, the generation performance consistently improves as the ETR increases. This correlation is expected, as a higher tolerance for error typically corresponds to higher accuracy. It is worth noting that when ETR is 0.10, the accuracy with all three fragment percentages exceeds 95\%, highlighting \textit{RM-Gen}'s efficiency in generating precise radio maps even with just 5\% measured RSS data. 

Contrastingly, in Fig.~\ref{fig5a}(b), illustrating the generation performance in outdoor cellular network scenarios, while the model exhibits a similar trend, its performance is comparatively lower. For instance, at ETR = 0.10, the accuracy is around 90\%, which is over 5\% lower than the evaluated cases in Fig.~\ref{fig5a}(a). This difference can be attributed to the complexity of scenarios as well as the inherent characteristics of mmWave and sub-6GHz signals. Outdoor network settings are typically larger than indoor scenarios and include more complex elements like buildings. Additionally, sub-6GHz signals, commonly used in outdoor environments, penetrate obstacles more effectively, resulting in more irregular RSS distributions and posing a greater challenge for accurate generation.

\subsubsection{RSS distribution comparison}

\begin{figure}[htbp]
\centerline{\includegraphics[width=0.75\linewidth]{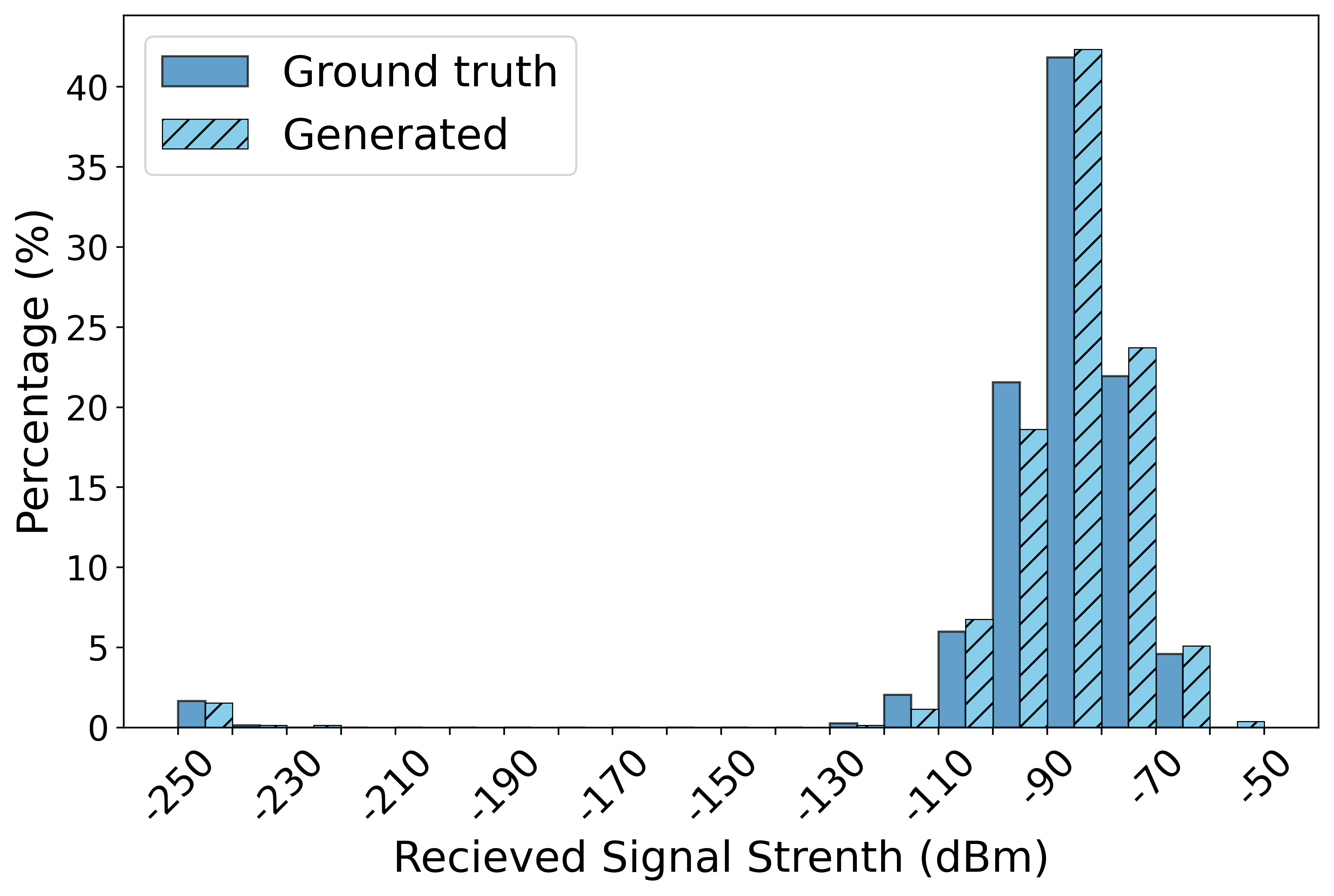}}
\caption{RSS distribution for indoor mmWave WLANs.}
\label{fig6}
\end{figure}

\begin{figure}[htb]
\centerline{\includegraphics[width=0.75\linewidth]{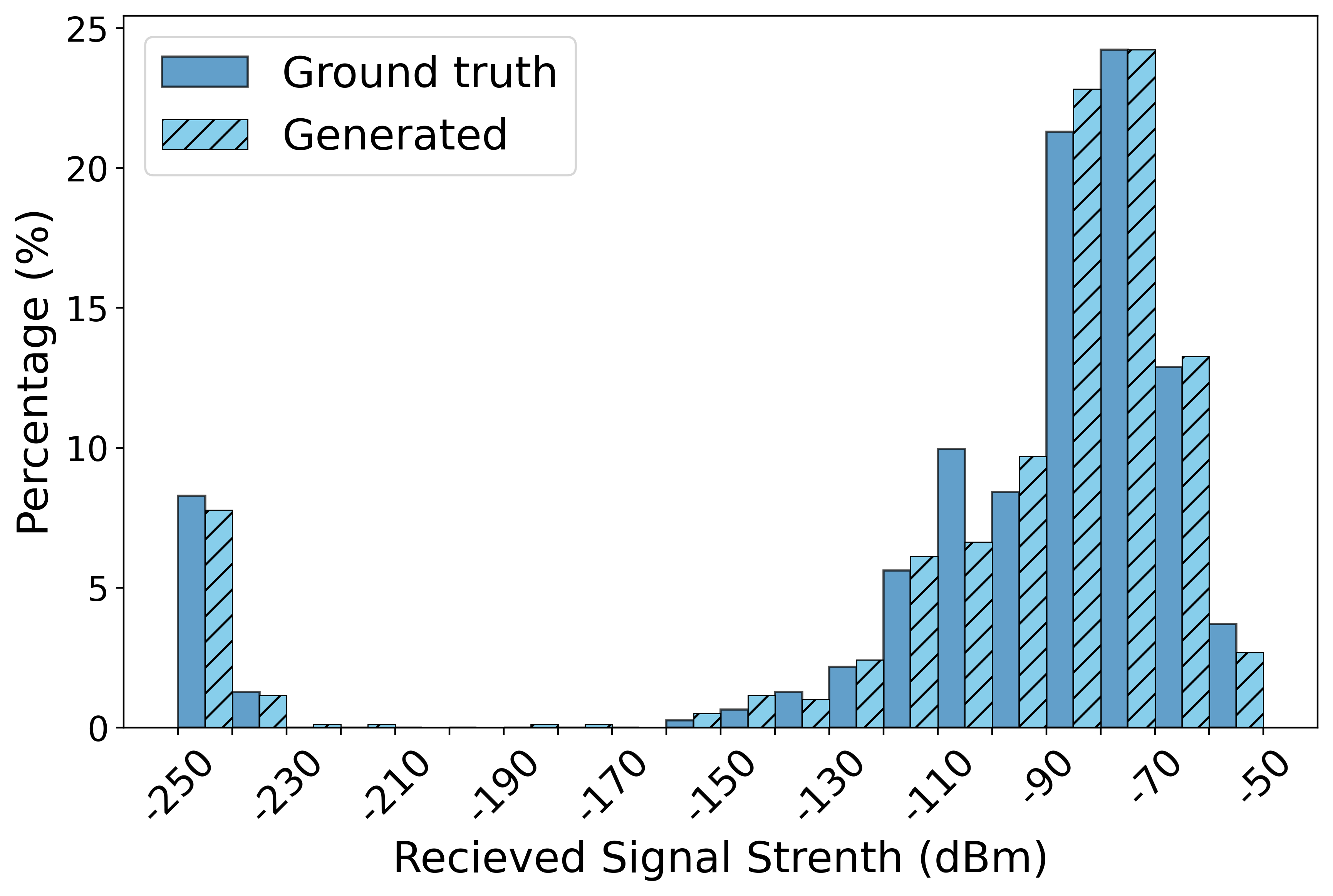}}
\caption{RSS distribution for outdoor sub-6GHz networks.}
\label{fig7}
\end{figure}
To further explore such differences between mmWave and sub-6GHz signals, we randomly select two samples from the indoor and outdoor radio maps generated by \textit{RM-Gen} and zoom in on the data results. Fig.~\ref{fig6} and Fig.~\ref{fig7} depict the RSS distribution between the generated radio maps and the ground truths for these samples. As seen in Fig.~\ref{fig6}, the RSS values in indoor mmWave scenarios cover a wide range from -250 dBm to -50 dBm, with a concentration between -100 dBm and -70 dBm. Notably, the RSS between -80 dBm and -90 dBm accounts for the highest proportion with a clear peak, representing more than 40\% of the RSS distribution.

Conversely, Fig.~\ref{fig7} illustrates that outdoor cases, while covering a similar range, exhibit a more uniform distribution, primarily between -120 dBm and -60 dBm. In particular, the peaks for outdoor sub-6GHz RSS, situated between -90dBm and -70dBm, are more dispersed than indoor mmWave RSS, reflecting the variability and instability of the outdoor radio profiles. {From Fig.~\ref{fig7}, it can be observed that outdoor scenarios give more distribution values in lower received signal strengths due to the presence of more physical obstacles, such as buildings and larger objects, which significantly weaken the RSS in the blocked areas.}

Besides these observations that underscore the distinct propagation characteristics of indoor mmWave and outdoor sub-6GHz signals, the results reveal that the RSS distribution of our generated samples closely aligns with the ground-truth results, affirming the \textit{RM-Gen}'s capability to accurately synthesize radio maps in varying scenarios. 

\begin{figure}[t]
\centerline{\includegraphics[width=1\linewidth]{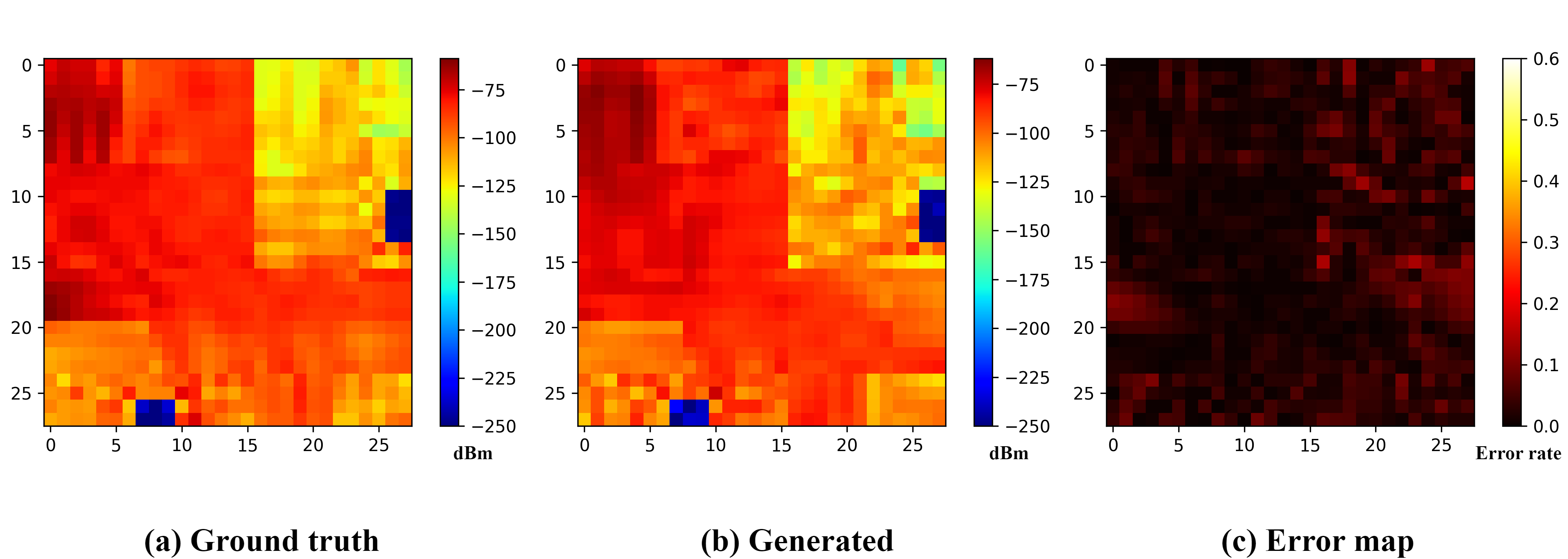}}
\caption{Visualization of generated radio maps for indoor cases.}
\label{fig8}
\end{figure}

\begin{figure}[t]
\centerline{\includegraphics[width=1\linewidth]{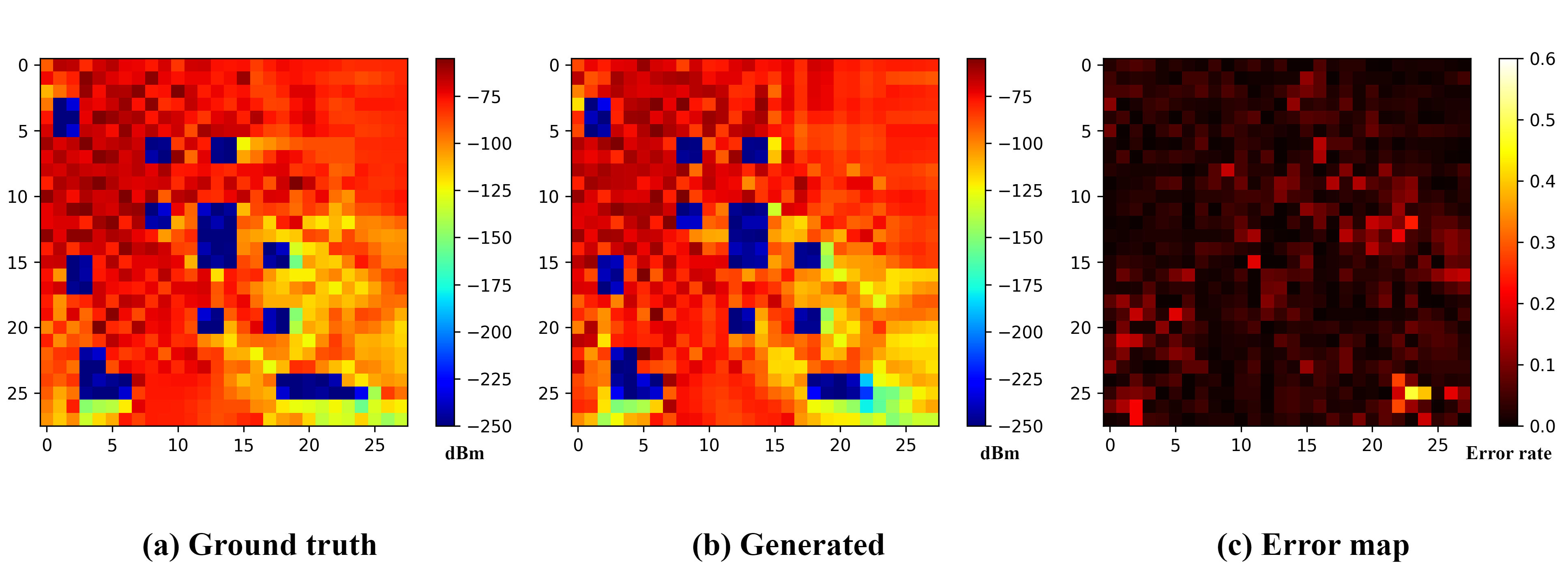}}
\caption{Visualization of generated radio maps for outdoor cases.}
\label{fig9}
\end{figure}
\subsubsection{Visual representations}

We also provide visual representations of the generation results from \textit{RM-Gen}. Specifically, Figs.~\ref{fig8}(a)-(b) show a ground-truth radio map and a corresponding generated radio map with the 10\% RSS fragments, respectively, while Fig.~\ref{fig8}(c) illustrates the error map between them. 
When considering the environment configuration depicted in Fig.~\ref{fig3}(a), it is notable that both radio maps shown in Figs.~\ref{fig8}(a)-(b) exhibit similar trends when examining each isolated area within the scenario. Moreover, Fig.~\ref{fig8}(c) indicates that the error rate in most regions is less than 0.1, affirming the accuracy of generated radio map.

Likewise, Fig.~\ref{fig9} shows the generation results for an outdoor network scenario, demonstrating a high alignment between the generated radio map and the ground-truth radio map. However, upon comparing the error maps from Fig.~\ref{fig8}(c) and Fig.~\ref{fig9}(c), it is evident that the outdoor scenario exhibits more ``polluted'' points with higher error rates. This finding is consistent with our previous evaluation results that \textit{RM-Gen} performs better in generating indoor radio maps.

\subsection{Radio Map Generation with Tx Locations}

\begin{figure}[t]
\centerline{\includegraphics[width=0.8\linewidth]{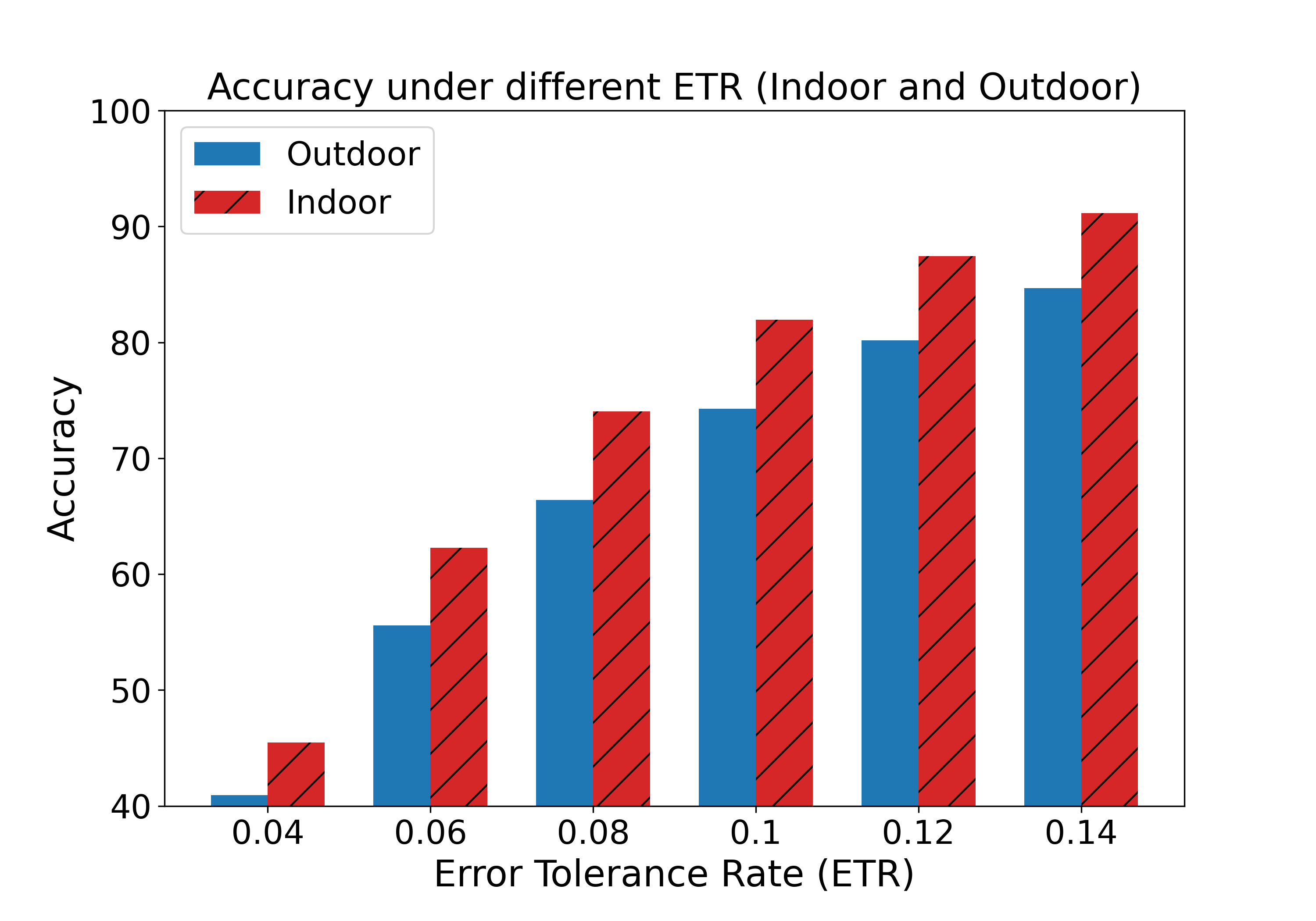}}
\caption{Generation accuracy using Tx locations as condition.}
\label{fig10}
\end{figure}

In addition to using partial RSS fragments, we investigate the feasibility of employing only Tx locations as input conditions to \textit{RM-Gen} in radio map generation. The conditional diffusion model is trained with two Tx locations as input on the RM-In and RM-Out datasets, and the results reported in Fig.~\ref{fig10}.
It is observed that as the ETR increases, the generation accuracy improves in both indoor and outdoor network scenarios. Notably, the model performs significantly better in indoor scenarios than in outdoor scenarios. This result is consistent with our observation in Sec.~V.B. 
Particularly, when comparing the generation accuracy depicted in Fig.~\ref{fig5a}, the results are notably inferior when Tx locations are employed as conditions compared to RSS fragments. Specifically, at ETR=0.10, it achieves accuracies of only 81.95\% and 74.26\% in indoor and outdoor network scenarios, respectively. 

The reduced accuracy observed when employing Tx locations as conditions is attributed to the limited information they convey compared to RSS fragments.  While RSS fragments offer rich local features of signal distributions, Tx location data primarily provides signaling source information, posing challenges for accurately reconstructing full radio maps.
However, despite these limitations, utilizing Tx location as conditions presents intriguing practical potential. Generating radio maps based on planned BS/AP positions can be particularly advantageous in contexts where performing measurements or ray-tracing analysis is impractical or costly. Future research could focus on enhancing the model's ability to interpret and effectively utilize Tx location information alongside sparse RSS measurements, potentially opening up new avenues for efficient and accurate radio map generation in these advanced network scenarios.

\subsection{Results of Comparison With Baselines}
{Before the rise of diffusion models, GAN-based methods were among the most advanced and widely used approaches for image generation. In light of their significance, we compare our \textit{RM-Gen} model with two well-established GAN-based models: a conditional GAN (cGAN) \cite{mirza2014conditional} and a pix2pix model \cite{isola2017image}.}
The GAN framework~\cite{creswell2018generative} comprises two neural networks, a generator $G$ and a discriminator $D$, that compete in a game-theoretic scenario. 
Expanding on the foundational GAN concept, cGAN incorporating additional conditional information into the generation process, aligning closely with the requirements of our specific problem. 
Similarly, the pix2pix model enhances the cGAN by integrating an L1 loss term, which ensures pixel-wise accuracy between the generated and target images.

\begin{itemize}
    \item \textbf{\textit{cGAN}}: The training process of cGAN begins by sampling a noise vector $z$ from a noise prior distribution $p_g(z)$, which serves as the initial seed for data generation. During each iteration of training, the discriminator is trained first and receives a batch of real data samples $x$ from the dataset as well as a batch of generated data produced by the generator from the noise samples. We then update this discriminator by ascending its stochastic gradient as: $\nabla_{\theta_d} \frac{1}{m} \sum_{i=1}^m \big[\log D\left(x^{(i)} \mid e_{\theta}(c) \right) + \log \left(1 - D\left(G\left(z^{(i)} \mid e_{\theta}(c)\right)\right)\right) \big],$
    where $m$ is the batch size set as 64.
    Subsequently, the generator is updated by descending its stochastic gradient as: $\nabla_{\theta_g} \frac{1}{m} \sum_{i=1}^m \log \left(1 - D\left(G\left(z^{(i)} \mid e_{\theta}(c)\right)\right)\right).$

    \item {\textbf{\textit{pix2pix}}: The pix2pix model builds upon cGAN by introducing an additional L1 loss, which measures the pixel-wise difference between the generated output and the ground truth, helping to produce more accurate and realistic outputs. The objective of pix2pix can be written as a combination of the adversarial loss from cGAN and the L1 loss as: $\mathcal{L}_{\text{pix2pix}} = \mathcal{L}_{\text{cGAN}} + \lambda \mathcal{L}_{\text{L1}}$, where $\mathcal{L}_{\text{cGAN}}$ is the adversarial loss as defined earlier, and $\mathcal{L}_{\text{L1}} = \frac{1}{m} \sum_{i=1}^m \| x^{(i)} - G(x^{(i)} \mid e_{\theta}(c)) \|_1$. The parameter $\lambda$ is used to control the trade-off between the adversarial loss and the L1 loss. Unlike the original pix2pix, which uses PatchGAN as the discriminator \cite{isola2017image}, we employ an MLP-based discriminator in this baseline model.}

\end{itemize}

For handling conditional inputs in cGAN and pix2pix, we employs the same encoder architecture used in our \textit{RM-Gen} to extract features from the conditions. This ensures a consistent approach to feature extraction across different generative models, allowing for a fair comparison of their performance in generating radio maps under similar input conditions.

\begin{table}[htbp]
\caption{Performance comparison with baselines using RSS fragments as conditions.} 
\centering
\begin{tabular}{|c|c|c|c|c|c|c|}
\hline
\multicolumn{1}{|c|}{Dataset} & \multicolumn{3}{c|}{\textbf{RM-In}} & \multicolumn{3}{c|}{\textbf{RM-Out}} \\ \hline
Fragments (\%) & 5 & 10 & 15 & 5 & 10 & 15 \\ \hline
cGAN \cite{mirza2014conditional} & 67.03 & 69.28 & 74.24 & 60.19 & 62.51 & 65.80 \\ \hline
pix2pix \cite{isola2017image} & 86.67 & 89.72 & 90.88 & 71.93 & 75.15 & 79.30 \\ \hline
\textbf{\textit{RM-Gen}} & \textbf{96.85} & \textbf{97.78} & \textbf{98.09} & \textbf{89.54} & \textbf{89.95} & \textbf{90.97} \\ \hline
\end{tabular}
\label{fragment}
\end{table}

The experimental results comparing the performance of our \textit{RM-Gen} with cGAN and pix2pix are summarized in Table \ref{fragment} and Table \ref{location}. The evaluation is conducted using both two types of conditions, i.e. RSS fragments and Tx locations.
First, Table \ref{fragment} shows the accuracy of the radio maps generated by cGAN, pix2pix, and \textit{RM-Gen} using RSS fragments as conditions with ETR of 0.10. It is evident that \textit{RM-Gen} significantly outperforms cGAN by more than 20\% and pix2pix by around 10\% across various fragment percentages and both indoor and out door scenarios. {We also observe that the pix2pix performs better than cGAN since the addition of the L1 loss ensures that the generated radio maps align more closely with the ground-truth data.  However, despite this improvement, pix2pix is still less effective than \textit{RM-Gen}, likely because our diffusion model iteratively refines data through denoising steps during the generation process, enabling it to better capture and reconstruct fine-grained signal variations in complex environments.} Notably, in the RM-In dataset using RSS fragments as conditions, \textit{RM-Gen} attains an accuracy of 96.85\% with just 5\% data fragments, whereas cGAN achieves only 67.03\% accuracy under the same fragment percentage. This substantial improvement highlights the effectiveness of \textit{RM-Gen} in processing and leveraging small data pieces to generate complete radio maps in various network scenarios.
Table \ref{location} demonstrates the accuracy of these three models when using Tx locations as conditions with the ETR range from 0.10 to 0.14. Similarly, \textit{RM-Gen} shows superior performance compared with cGAN and pix2pix across all error tolerance rates and both network scenarios. 
These results suggest that \textit{RM-Gen} is more effective in capturing underlying spatial information provided by the AP locations for generating accurate radio maps.

\begin{table}[htbp]
\caption{Performance comparison with baselines using Tx locations as conditions.} 
\centering
\begin{tabular}{|c|c|c|c|c|c|c|}
\hline
\multicolumn{1}{|c|}{Dataset} & \multicolumn{3}{c|}{\textbf{RM-In}} & \multicolumn{3}{c|}{\textbf{RM-Out}} \\ \hline
ETR & 0.10 & 0.12 & 0.14 & 0.10 & 0.12 & 0.14 \\ \hline
cGAN \cite{mirza2014conditional} & 74.53 & 81.91 & 87.33 & 69.71 & 75.46 & 79.75 \\ \hline
pix2pix \cite{isola2017image} & 79.47 & 85.09 & 88.93 & 70.02 & 76.85 & 82.08 \\ \hline
\textbf{\textit{RM-Gen}} & \textbf{81.95} & \textbf{87.44} & \textbf{91.14} &\textbf{ 74.26} & \textbf{80.16} & \textbf{84.69} \\ \hline

\end{tabular}
\label{location}
\end{table}

Interestingly, our results indicated that cGAN and pix2pix perform relatively better when using Tx locations as conditions compared to RSS fragments. 
This could be attributed to the fact that Tx locations offer a more structured and less complex form of data compared to RSS fragments, as GAN-based models directly learn the data distribution through adversarial training between a generator and a discriminator, which often leads to instability issues, such as mode collapse, gradient vanishing, and exploding gradients, making it difficult for GAN-based models to effectively capture complex data patterns like RSS fragments. In contrast, our diffusion-based model adopts a gradual denoising process, starting from Gaussian noise and iteratively refining the sample towards the data distribution. This allows for a smoother and more stable learning trajectory, as each step involves learning a simpler transformation. From this perspective, \textit{RM-Gen} offers greater stability and transparency, which can be advantageous in learning and exploiting complex data distributions for radio map generation.

\subsection{Practical Use Cases of RM-Gen}

\begin{figure}[t]
\centerline{\includegraphics[width=1\linewidth]{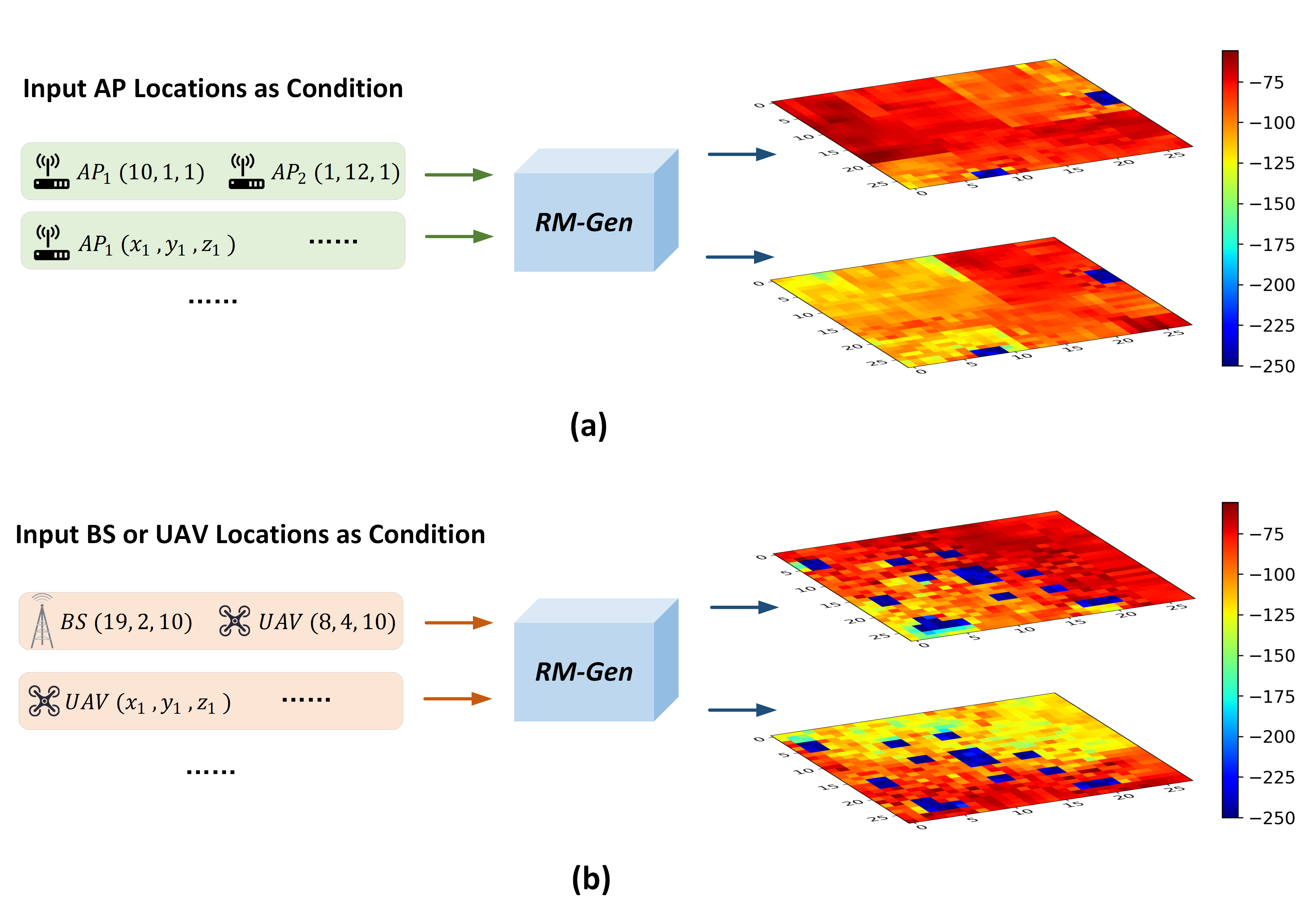}}
\caption{Use cases of our \textit{RM-Gen}.}
\label{fig11}
\end{figure}

Lastly, we showcase some practical use of our \textit{RM-Gen} in downstream network tasks. In the indoor scenario, \textit{RM-Gen} can demonstrate its capability by generating radio maps using wireless APs as the Tx sources, as shown in Fig.~\ref{fig11}(a). This application finds relevance in settings like office buildings, shopping malls, and large residential areas. The generated radio map efficiently illustrates signal coverage and strength across the indoor space, originating from strategically positioned APs, without the need for prior measurements. This visualization facilitates the identification of potential areas of weak signal strength and enables network administrators to optimize AP placement and configuration.

In outdoor scenarios, \textit{RM-Gen} extends its application to encompass more expansive and dynamic environments by incorporating BSs and UAVs as Tx sources, as depicted in Fig. 11(b). 
The model can generate radio maps predicting the dynamic coverage area provided by BSs and UAVs, which is crucial for backhaul planning and optimizing the trajectory of UAVs to ensure wide-area coverage and robust connectivity. 
The use of UAVs as mobile AP adds a layer of flexibility, enabling real-time compensation to the network coverage holes in reaction to emergency situations, thereby significantly enhancing the resilience and adaptability of wireless networks.

\vspace{+0.2cm}
\section{Conclusion}

In this paper, we presented \textit{RM-Gen}, a novel framework leveraging conditional diffusion models for radio map generation in wireless networks. Our approach addresses the challenges of estimating accurate radio maps with minimal prior knowledge by utilizing sparse RSS fragments and transmitter locations. We introduced an environment-aware method that strategically selects RSS fragments based on the quantified environmental complexity, which was shown to significantly enhance the model's performance when expert knowledge is not available. 
Through comprehensive evaluations, we investigated the impact of utilizing partial RSS fragments and Tx locations as conditions for map generation. The results showcased the remarkable accuracy of the generated radio maps, achieving accuracy rates of 95\% for indoor WLANs and 90\% for outdoor cellular networks. 
Besides, \textit{RM-Gen} consistently outperformed the conventional cGAN and pix2pix approaches in all experimental settings. The results underscore the potential of \textit{RM-Gen} in practical applications, such as network planning and optimization, by providing a cost-effective and adaptable solution for radio map generation. 
Future work could be focused on enhancing the generative model's capability to handle more complex and dynamic network environments, potentially integrating additional side-user information to further improve its applicability.

\section*{Acknowledgment}
This research was supported by the National Science Foundation through Awards CNS--2312138, SaTC--2350075, ECCS-2434054, and CNS--2312139.

\bibliographystyle{IEEEtran}
\bibliography{reference}

\end{document}